\documentclass[12pt,preprint]{aastex}

\usepackage{natbib}
\def\mum{${\rm \mu m}$}
\def\sta{C$_{24}$H$_{12}$}
\def\stb{C$_{54}$H$_{18}$}
\def\stc{C$_{66}$H$_{20}$}
\def\std{C$_{78}$H$_{22}$}
\def\ste{C$_{96}$H$_{24}$}
\def\stf{C$_{110}$H$_{26}$}
\def\stg{C$_{112}$H$_{26}$}
\def\sth{C$_{130}$H$_{28}$}

\shorttitle{Large symmetric PAHs}
\shortauthors{Bauschlicher et al.}
\begin{document}
\title { The infrared spectra of very large, compact, highly symmetric, \\
polycyclic aromatic hydrocarbons (PAHs)}
\author{Charles W. Bauschlicher, Jr.\altaffilmark{1}, 
Els Peeters\altaffilmark{2,3,4}, Louis J. Allamandola\altaffilmark{5}}

\altaffiltext{1}{NASA-Ames Research Center, Space Technology Division,
Mail Stop 230-3, Moffett Field, CA 94035, USA;
Charles.W.Bauschlicher@nasa.gov}

\altaffiltext{2}{Department of Physics and Astronomy, PAB213,
University of Western Ontario, London, ON N6A 3K7, Canada;
epeeters@uwo.ca}

\altaffiltext{3}{SETI Institute, 515 N. Whisman Road, Mountain View,
CA94043, USA}

\altaffiltext{4}{NASA-Ames Research Center, Space Science Division,
Mail Stop 245-6, Moffett Field, CA 94035, USA}

\altaffiltext{5}{NASA-Ames Research Center, Space Science Division,
Mail Stop 245-6, Moffett Field, CA 94035, USA;
Louis.J.Allamandola@nasa.gov}


\begin{abstract}

The mid-infrared spectra of large PAHs ranging from \stb\, to \sth\,
are determined computationally using Density Functional Theory.
Trends in the band positions and intensities as a function of PAH
size, charge and geometry are discussed.  Regarding the 3.3, 6.3 and
11.2 \mum\, bands similar conclusions hold as with small PAHs.

This does not hold for the other features. The larger PAH cations and
anions produce bands at 7.8 \mum\, and, as PAH sizes increases, a band
near 8.5 \mum\, becomes prominent and shifts slightly to the red.  In
addition, the average anion peak falls slightly to the red of the
average cation peak. The similarity in behavior of the 7.8 and 8.6
\mum\, bands with the astronomical observations suggests that they
arise from large, cationic and anionic PAHs, with the specific peak
position and profile reflecting the PAH cation to anion concentration
ratio and relative intensities of PAH size. Hence, the broad astronomical
7.7 \mum\, band is produced by a mixture of small and large PAH
cations and anions, with small and large PAHs contributing more to the
7.6 and 7.8 \mum\, component respectively.

For the CH out-of-plane vibrations,  the duo hydrogens couple with the
solo vibrations and produce bands that fall at wavelengths slightly
different than their counterparts in smaller PAHs. As a consequence,
previously deduced PAH structures are altered in favor of more compact
and symmetric forms. In addition, the overlap between the duo and trio
bands may reproduce the blue-shaded 12.8 \mum\, profile. 

\end{abstract}

\section{Introduction}

\begin{figure*}
\plotone{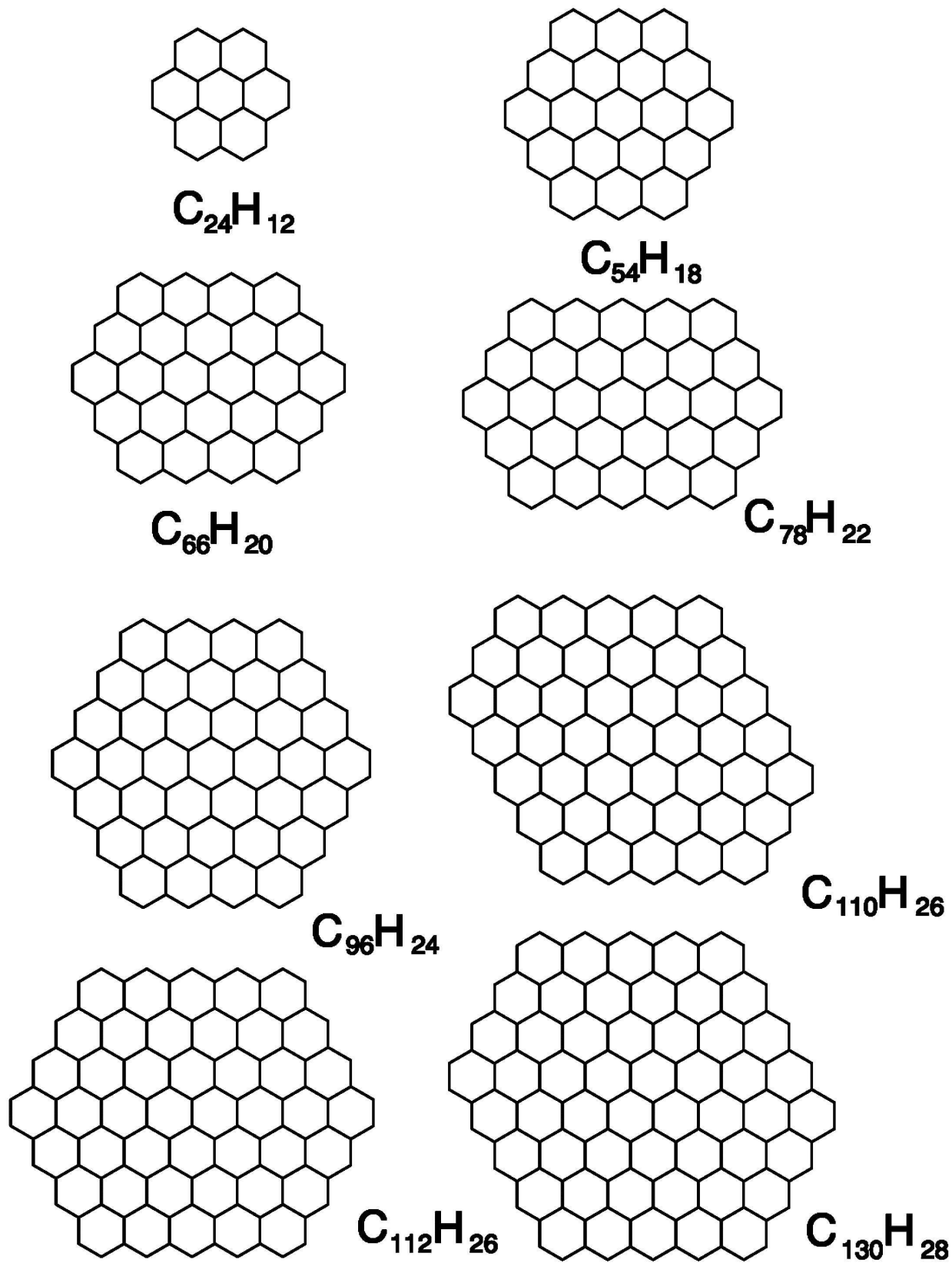}
\caption{The structure of the PAHs studied in this paper.}
\label{structure}
\end{figure*}

Polycyclic aromatic hydrocarbons (PAHs) are now thought to be the most
plentiful and widespread class of organic compounds in the
universe. Their infrared (IR) signature is associated with many
different galactic and extragalactic objects \citep[e.g.][]{isoparijs,
Helou:normalgal:00, Vermeij:pahs:01,Peeters:colorado, Bregman:05,
Brandl:06, Sellgren:07, SmithJD:07, Galliano:07}, and they account for
10 to 30\% of cosmic carbon \citep[][]{Snow:c:95, Puget:revpah:89,
ATB, Li:01, Draine:07}.  For most objects in which the PAH features
are present, their spectroscopic signature dominates the mid-IR
emission \citep[e.g.][]{Uchida:RN:00,Verstraete:prof:01,
Peeters:cataloog:02, Peeters:colorado, Onaka:colorado,Sloan:05,
Brandl:06}.  It is well established that ionization of PAH molecules
dramatically enhances the emission in the 6-9 \mum\, region,
suggesting that most of the observed IR emission arises from highly
vibrationally excited cations.  This excitation generally proceeds via
the absorption of ultraviolet (UV), visible (Vis), and near-IR (NIR)
radiation \citep[e.g.][]{Li:02, Li:04, Mattioda:model:05}.  While the
early, low to moderate resolution spectroscopic observations suggested
that the dominant PAH features are invariant, analysis of higher
resolution spectra of many objects revealed that the PAH features vary
between different classes of object and spatially within extended
objects \citep[e.g.][]{Peeters:prof6:02, vanDiedenhoven:chvscc:03,
Joblin:isobeyondthepeaks:00, Bregman:05, Sloan:05, Compiegne:07},
showing that the details in the emission spectrum depend on, and
therefore reflect, the specific PAH molecules present and the
conditions within the emission zones.

Most previous work on the IR spectroscopic properties of PAHs focused
on species containing about 50 or fewer carbon atoms because large
PAHs are not readily available for experimental study and
computational techniques for such large systems were not practical.
While large PAH accessibility remains limited, computational
capabilities have increased and the spectra of large PAHs can be
determined with good precision.  Here we report and discuss the
computational IR spectra of several large, compact, and symmetric
PAHs, namely \stc, \std, \stf, \stg\, and \sth.  When combined with our
previous work on the large PAHs \stb\, and \ste\,
\citep{Bauschlicher:C54:00, Bauschlicher:c96:02}, these spectra
provide new insight into the effect of PAH size and structure on their
IR spectra.  The IR spectra of comparably sized, but less compact and
less symmetric, PAHs are discussed in a subsequent paper (Bauschlicher
et al. in preparation, paper II).  These two studies, together with
the previous work on nitrogen substitution, substantially deepen our
understanding of the PAH populations that contribute to the
astronomical spectra.
 
This work is presented as follows.  The computational methods used are
described in Sect. \ref{method} and the spectra are presented and
discussed in Sect. \ref{lab}.  Applications to the astronomical
observations and conclusions regarding the astronomical PAH population
are given in Sect. \ref{astro}. The paper is concluded in
Sect. \ref{con}.

\section{Model and Methods}
\label{method}

The PAHs considered here are shown in Fig.~\ref{structure}, along with
the previously studied coronene (\sta), circumcoronene (\stb), and
circumcircumcoronene (\ste) molecules.  The spectra for the neutral,
cation and anion forms for all these PAHs have been computed.  The
geometries are optimized and the harmonic frequencies and IR
intensities are computed using the B3LYP \citep{Stephens:94} hybrid
\citep{Becke:93} functional in conjunction with the 4-31G basis sets
\citep{Frisch:84}.  The calculations are performed using the Gaussian
03 computer codes \citep{Frisch:03}.

The number of bands determined for these large species is so great
that we do not show all of the data here.  These data are available
upon request from the authors and will become part of the publicly
available Ames PAH IR Spectral Database which is now under
construction.  To illustrate these results, synthetic spectra are
presented in which the computed frequencies have been scaled by 0.958
and the behavior of some of the more important band positions are
discussed.  The scaling factor of 0.958 has been found to bring the
computationally determined PAH vibrational frequencies into very good
agreement with experimentally measured spectra
\citep{Langhoff:neutionanion:96, Bauschlicher:97}.  For example, most
computational and experimental peak positions fall within 5 cm$^{-1}$
of each other, some within 10 cm$^{-1}$, and a handful within 15
cm$^{-1}$. The observed trends in peak position with size and
between cations, anions and neutrals should be more accurate than
this absolute uncertainty, hence the small shifts we report in this paper
should be valid.  The intensities are unscaled, despite the potential
factor of 2 overestimation of the computed intensity for the CH
stretching modes of the neutrals that has been discussed previously
\citep{Bauschlicher:97, Hudgins:spectrochimica:01}.  To permit
comparison of these absorption spectra with astronomical observations
which are measured in emission, the natural mid-IR linewidth for a
large molecule emitting under interstellar conditions has to be taken
into account.  Up to now, this has been taken as about 30 cm$^{-1}$
across the mid-IR.  As discussed in Cami et al. (in preparation), the
natural linewitdth can be band dependent.  Here a linewidth of 30
cm$^{-1}$ is taken for the bands shortward of 9 \mum, 10 cm$^{-1}$ for
the bands longward of 10 \mum, values consistent with current
observational and theoretical constraints. For the 9 to 10 \mum\,
region, the FWHM is scaled in a linear fashion (in wavenumber space)
from 30 to 10 cm$^{-1}$. In addition to ignoring any further
variations of linewidth as a function of mode, Fermi resonances are
not taken into account.  Despite these limitations, these idealized
spectra can be useful in better understanding the astronomical
spectra.

\section{Results and Discussion }
\label{lab}

\begin{table}
\caption{\label{t1}  The C-H stretching band position maximum ($\lambda$, in $\mu$m), total intensity (I), and
intensity per CH (I(CH)) for the PAHs shown in Fig. \ref{structure}.  The intensities are  in km/mol. }
\begin{center}
\begin{tabular}{l@{\hspace{9pt}}rrr@{\hspace{9pt}}rrr@{\hspace{9pt}}rrr}
\hline \\[-5pt]
Molecule & \multispan3 \hfil Cation \hfil &\multispan3 \hfil  Neutral \hfil &\multispan3 \hfil  Anion \hfil \\
 &  \multicolumn{1}{c}{$\lambda$} &   \multicolumn{1}{c}{I} &   \multicolumn{1}{c}{I(CH)}&  \multicolumn{1}{c}{$\lambda$} &  \multicolumn{1}{c}{I}&  \multicolumn{1}{c}{I(CH)} &   \multicolumn{1}{c}{$\lambda$} &  \multicolumn{1}{c}{I}&   \multicolumn{1}{c}{I(CH)}\\[5pt]
C$_{24}$H$_{12}$ &  3.238&    58&4.8 &3.262&    296&24.7& 3.296&    787& 65.6\\
C$_{54}$H$_{18}$ &  3.248&   286& 15.9&3.266&    577& 32.1& 3.283&   1078& 59.5\\
C$_{66}$H$_{20}$ &  3.250&   365&18.3&3.264&    687& 34.4& 3.280&   1208& 60.4\\
C$_{78}$H$_{22}$ &  3.252&   452& 20.6&3.264&    800& 36.4& 3.278&   1359& 61.8\\
C$_{96}$H$_{24}$ &  3.255&   631& 26.3&3.265&    931&38.8& 3.277&   1481& 61.7\\
C$_{110}$H$_{26}$ & 3.255&   731&28.1 &3.267&   1133&43.6& 3.277&   1811& 69.7\\
C$_{112}$H$_{26}$ & 3.256&   716& 27.5&3.266&   1070&41.2& 3.276&   1634& 62.9\\
C$_{130}$H$_{28}$ & 3.257&   871& 30.1&3.266&   1221&43.6& 3.276&   1846& 65.9\\[5pt]
\hline
\end{tabular}
\end{center}
\noindent
\end{table}

\begin{figure*}
   \begin{minipage}[c]{0.33\textwidth}
      \centering \includegraphics[width=\textwidth]{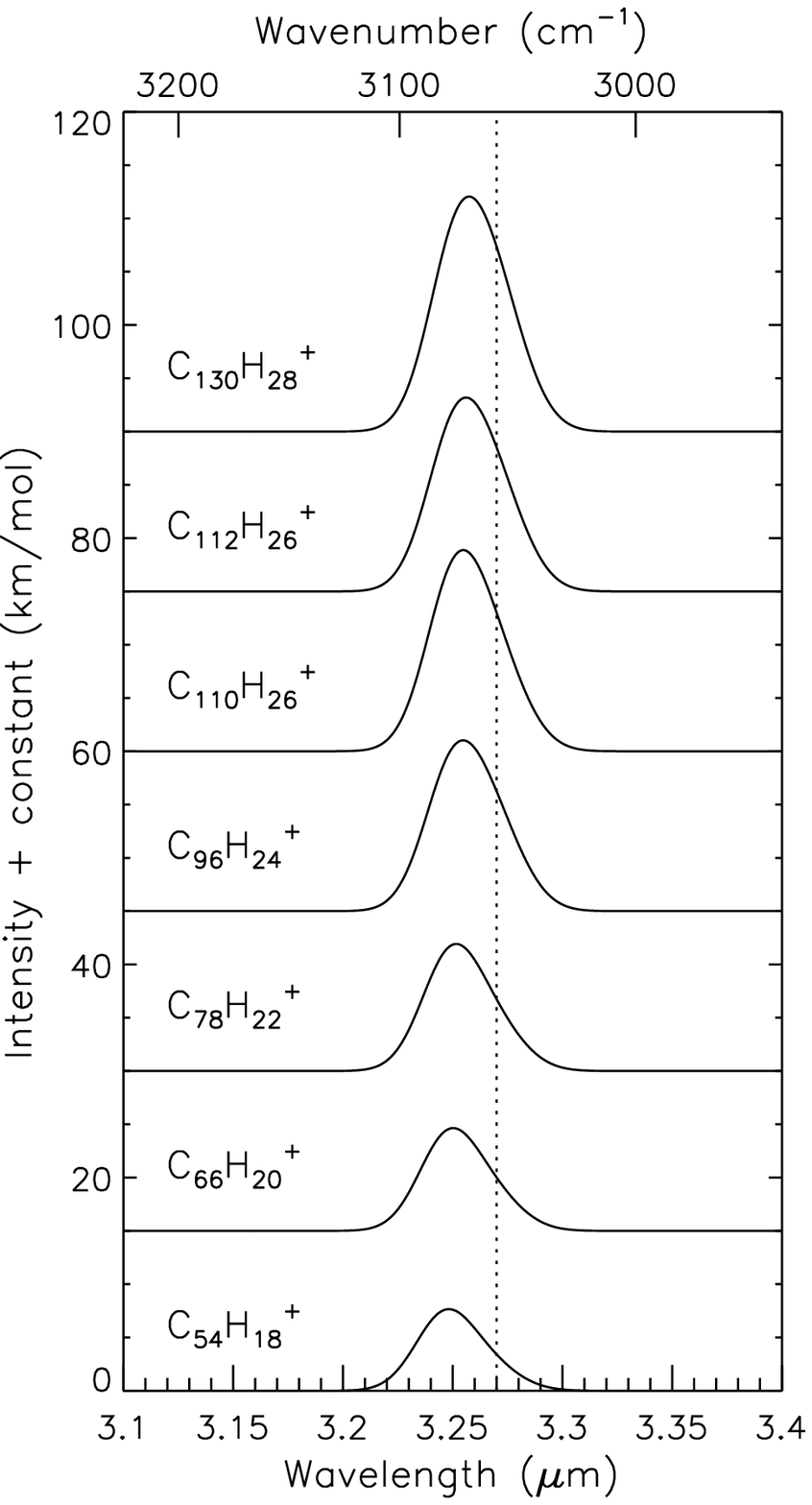}
   \end{minipage}
   \begin{minipage}[c]{0.33\textwidth}
      \centering \includegraphics[width=\textwidth]{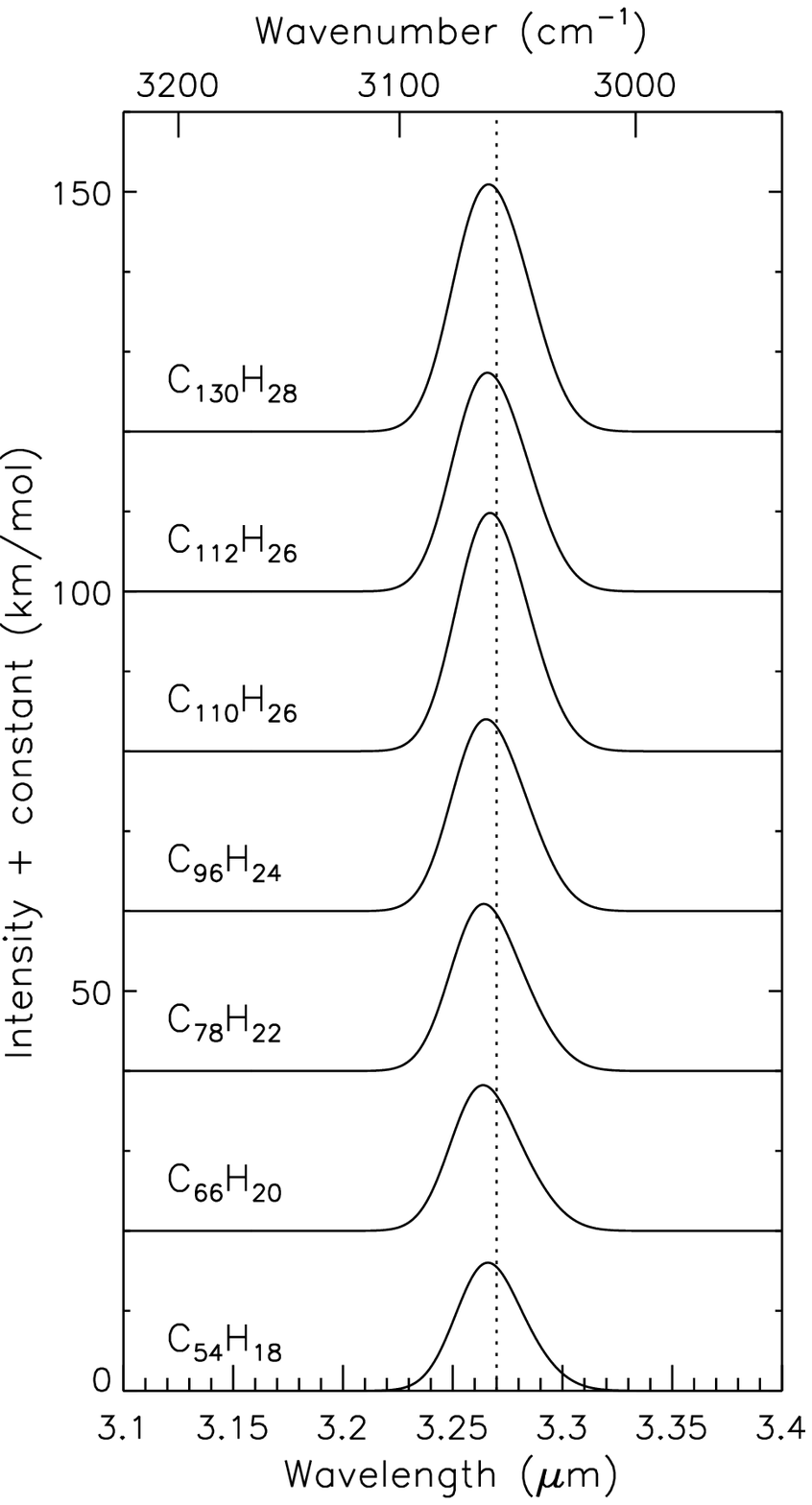}
   \end{minipage}
   \begin{minipage}[c]{0.33\textwidth}
      \centering \includegraphics[width=\textwidth]{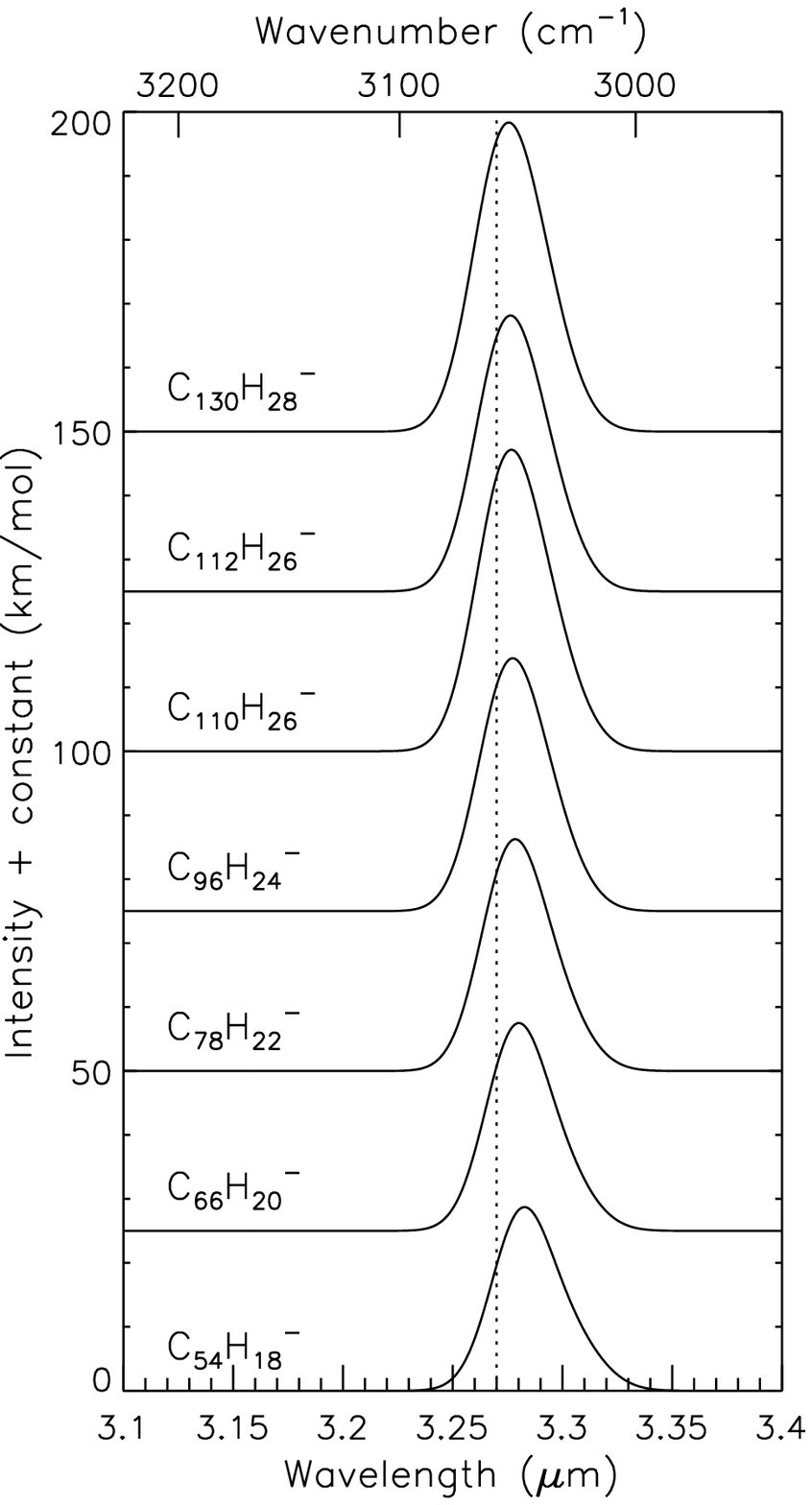}
   \end{minipage}
\caption{The synthetic absorption spectra in the 3 $\mu$m\, region for
the large symmetric PAH cations, neutrals and anions considered here. To
guide the eye, a dotted line at 3.27 $\mu$m\, is also shown.}
\label{fig_33}
\end{figure*}

\subsection{The C-H stretching vibrations (2.5 - 3.5 \mum).}
\label{lab_33}

The peak positions of the bands which dominate the CH stretching
region for each of the PAHs considered here are listed in Table
\ref{t1}. The spectra are shown in Fig.~\ref{fig_33}.  The peak
wavelength for the neutrals is essentially independent of PAH size and
falls at $\sim$3.26 \mum.  For both the cations and anions however,
there is a small shift with increasing size.  Including coronene
(\sta), the peak wavelength for the cations increases from 3.238 to
3.257 \mum\, with size while that for the anions drops from 3.296 to
3.276 \mum\, with PAH size.

Turning to band strengths, Table~\ref{t1} shows that the CH stretching
(CH$_{str}$) band intensities increase with PAH size for the neutrals,
cations, and anions.  While an increase in total band strength with
size is to be expected because the number of CH bonds increase with
PAH size, the non-additive behavior of the increase is noteworthy. To
illustrate this effect, Table~\ref{t1} also lists the integrated band
strengths (A values) per CH for each PAH.  Several important points
become apparent from this tabulation.  First, it is clear that the
behavior of coronene is not at all typical for the larger PAHs treated
here.  This is important because coronene has been considered
archtypical of astronomical PAHs, while significantly larger PAHs
dominate the cosmic mix.  Second, when one removes coronene from
further consideration, PAH CH$_{str}$ intensity per CH steadily increases
with PAH size for both the cations and neutrals, but much less so for
the anions.  Third, the CH$_{str}$ intensity drop upon ionization is small
for these large PAHs.  For example, the integrated intensity per CH
for the CH$_{str}$ in neutral \stg\, is 41 km/mol and it is 27 km/mol for
the cation form. In contrast, consider the A values for coronene in
Table~\ref{t1}.  Its A value is reduced from 24 km/mol to 5 km/mol,
nearly a factor of 5, upon ionization.  Significant CH$_{str}$ A value
reduction upon ionization is typical for all the small PAHs considered
to date \citep[e.g.][]{Langhoff:neutionanion:96, Bauschlicher:97,
Pathak:06, Malloci:07}.  Thus the dramatic reduction in the A value
per CH for the CH$_{str}$ upon ionization that is observed for small PAHs
disappears as PAH size increases.  This behavior is a consequence of
spreading the positive (or negative) charge over more carbon atoms
with increasing PAH size, thereby reducing the difference in the CH
stretching dipole derivative between the neutral, cation, and anion,
and hence reducing the difference in the observed intensities of the
CH stretching vibrations.

\begin{figure*}
   \begin{minipage}[c]{0.33\textwidth}
      \centering \includegraphics[width=\textwidth]{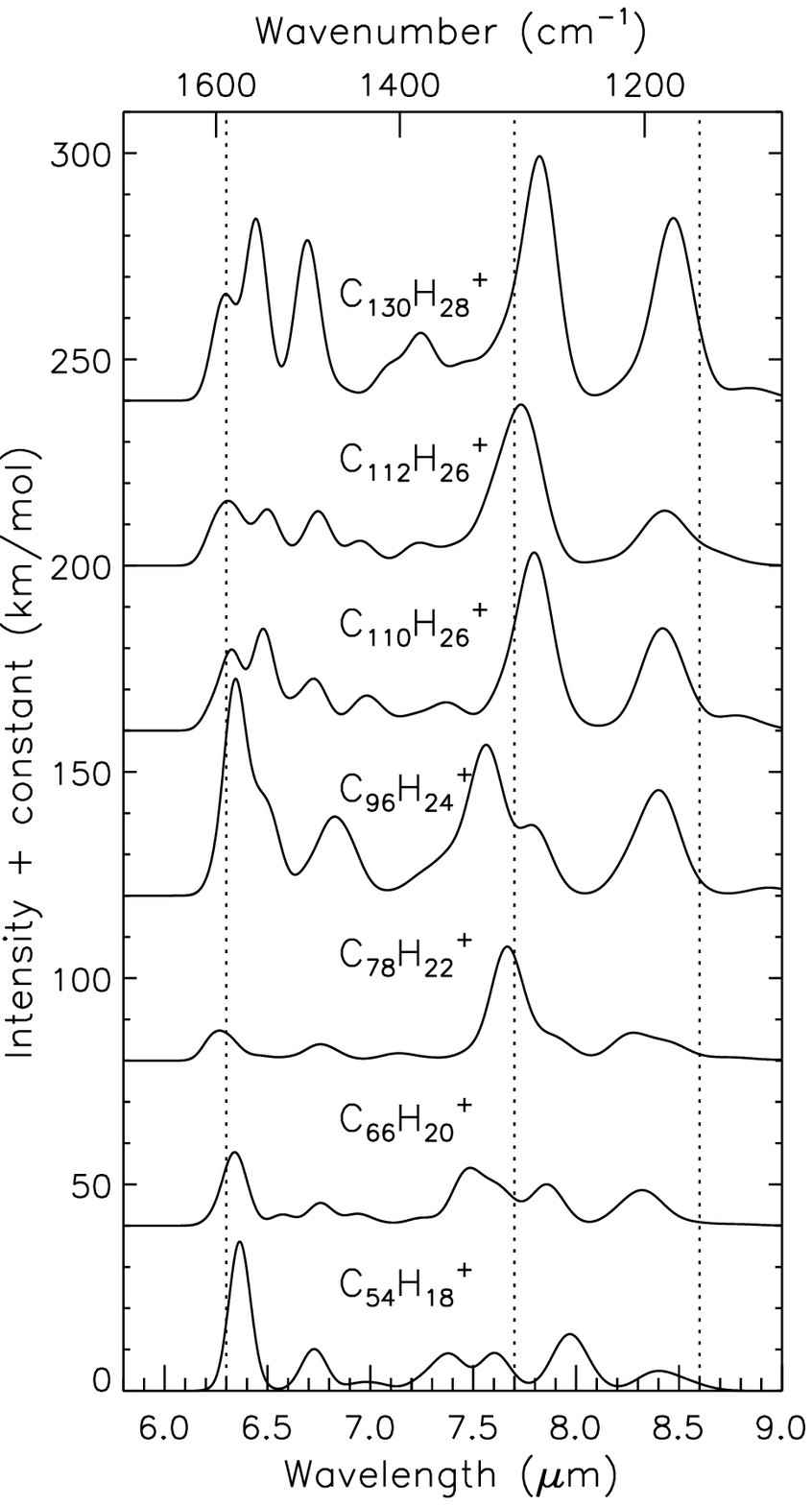}
   \end{minipage}
   \begin{minipage}[c]{0.33\textwidth}
      \centering \includegraphics[width=\textwidth]{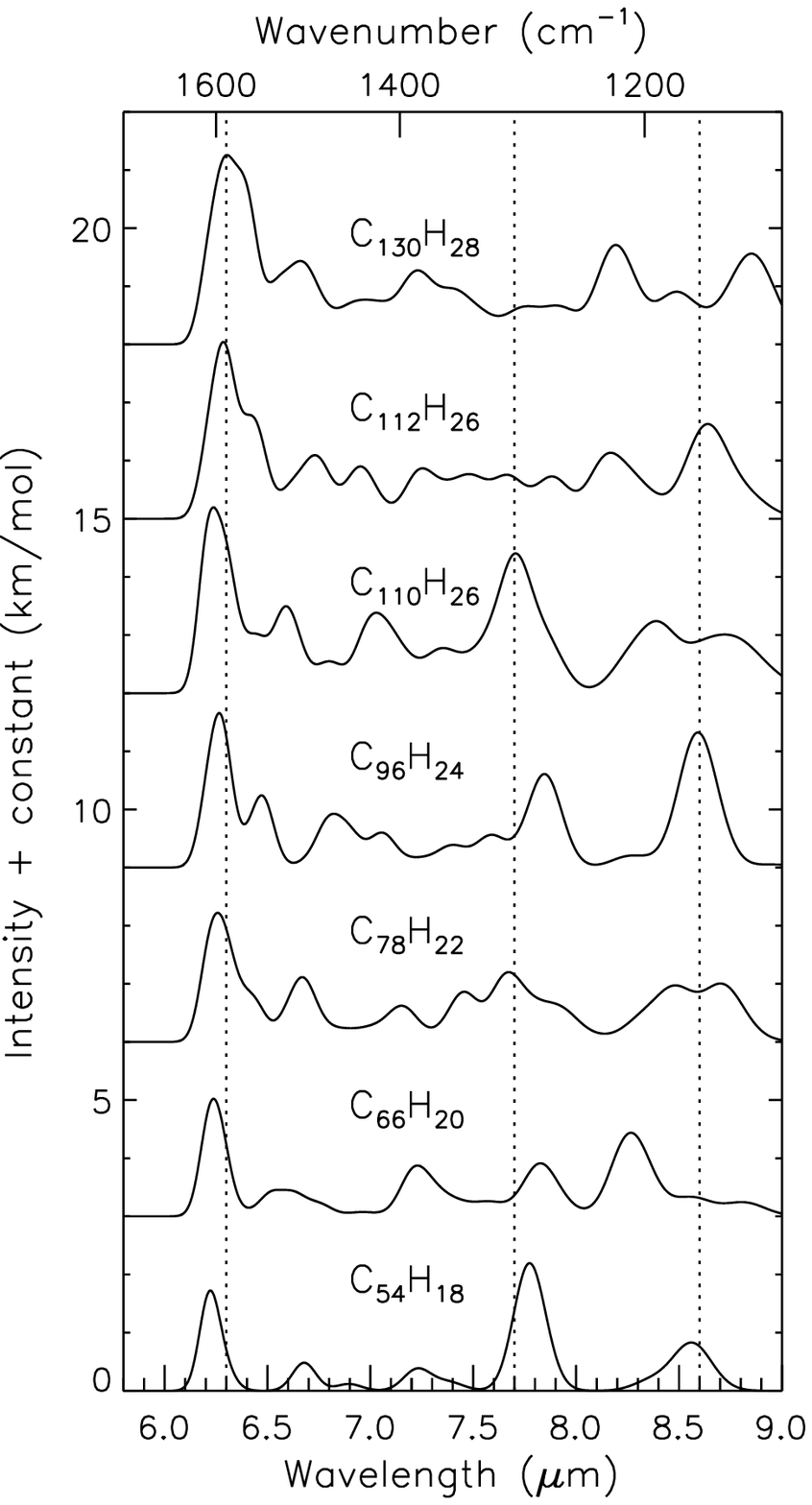}
   \end{minipage}
   \begin{minipage}[c]{0.33\textwidth}
      \centering \includegraphics[width=\textwidth]{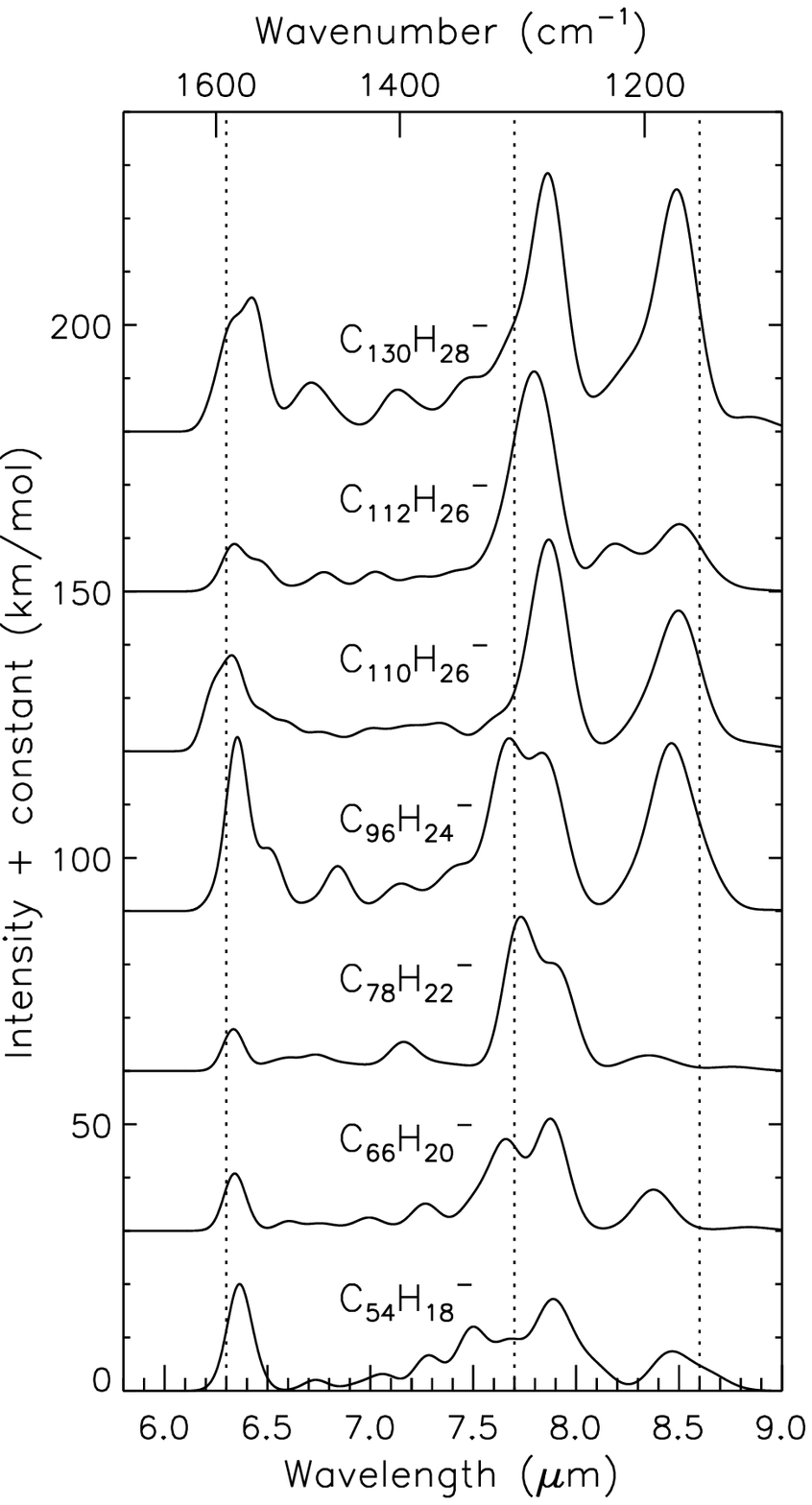}
   \end{minipage}
\caption{The synthetic absorption spectra in the 6 to 9 $\mu$m\, region for
the large symmetric PAH cations, neutrals and anions considered here. To
guide the eye, dotted lines at 6.3, 7.7 and 8.6 $\mu$m\, are also shown.}
\label{fig_62}
\end{figure*}

\subsection{The C-C Stretching and C-H in-plane bending vibrations (5 - 9 \mum).}
\label{lab_62}

\clearpage

\begin{deluxetable}{l@{\hspace{15pt}}rr@{\hspace{15pt}}rr@{\hspace{15pt}}rr}
\tablecaption{\label{t2} The 6-9 $\mu$m band position maxima ($\lambda$, in
$\mu$m) and total intensity (I).  The intensities are in km/mol. }
\tablehead{\colhead{Molecule} & \multicolumn{2}{l}{Cation} & \multicolumn{2}{l}{Neutral} & \multicolumn{2}{l}{Anion} \\
 &  \multicolumn{1}{c}{$\lambda$} &  \multicolumn{1}{c}{I} &  \multicolumn{1}{c}{$\lambda$} &  \multicolumn{1}{c}{I}&   \multicolumn{1}{c}{$\lambda$} &  \multicolumn{1}{c}{I}}
\startdata
C$_{24}$H$_{12}$ &      
    8.738&      53.8& 8.771&      13.1& 8.978&      71.6\\
   &  8.236&      93.2& 8.236&       2.0& 8.216&     131.3\\
    &     &          & 7.620&      48.1& 7.742&     410.1\\
  &   7.387&     435.3& 7.209&       1.6& 7.374&     118.1\\
   &       &          &     &           & 6.825&     152.3\\
  &   6.650&      54.1& 6.691&       2.7& 6.656&      23.7\\
  &   6.442&     465.9& 6.240&      26.3& 6.418&     269.2\\[5pt]
C$_{54}$H$_{18}$ &     8.403&     197.2& 8.559&      31.5& 8.467&     319.0\\
&     7.969&     477.6& & & 7.889&     801.4\\
&     7.603&     296.4& 7.774&      70.6& 7.686&     182.5\\
&     & & & & 7.501&     414.1\\
&     7.378&     369.6& 7.234&      16.5& 7.285&     202.7\\
&     6.980&      86.6& 6.894&       3.5& 7.059&     128.1\\
&     6.727&     357.4& 6.678&      15.6& 6.735&      70.1\\
&     6.365&    1189.3& 6.223&      61.1& 6.364&     778.2\\[5pt]
C$_{66}$H$_{20}$ &      & & 8.802&       6.9& 8.840&      23.4\\
 && & 8.547&       8.1\\
 &    8.320&     361.9& 8.267&      52.8& 8.376&     259.1\\
 &    7.857&     326.3& 7.827&      34.4& 7.874&     695.2\\
 &    7.486&     756.0& 7.567&       5.6& 7.658&     681.5\\
 &  & & 7.229&      41.9& 7.268&     174.4\\
 &    6.937&      97.2& 6.962&       1.8& 6.994&      93.5\\
 &    6.758&     195.7& & & 6.754&      45.4\\
 &    6.575&      77.5& 6.557&      31.9& 6.605&      60.7\\
 &    6.341&     692.7& 6.238&      77.8& 6.341&     365.7\\[5pt]
C$_{78}$H$_{22}$ &     & & 8.701&      32.4& 8.758&      26.1\\
&     8.279&     369.3& 8.482&      39.1& 8.354&     124.9\\
&     7.665&    1181.8& 7.672&      68.6& 7.732&    1497.7\\
& & & 7.457&      27.7\\
&     7.140&      85.4& 7.151&      30.7& 7.162&     261.2\\
&     6.757&     190.3& 6.669&      51.3& 6.733&     159.8\\
& & & & & 6.611&      83.7\\
&     6.267&     368.5& 6.258&     123.4& 6.335&     263.8\\ [5pt]
C$_{96}$H$_{24}$ &     8.938&      62.9\\
&     8.401&    1023.3& 8.592&      77.4& 8.463&    1352.4\\
&     & & 8.286&       4.7\\
&     7.782&     469.8& 7.846&      56.6& 7.833&     932.0\\
&     7.563&    1816.1& 7.590&      16.9& 7.675&    1451.7\\
&     & &7.404&      13.4\\
&     & &7.054&      20.5& 7.151&     194.5\\
&     6.827&     997.8& 6.823&      43.8& 6.841&     313.1\\
&      & & 6.471&      41.6& 6.497&     292.7\\
&     6.345&    2718.2& 6.265&     105.7& 6.354&    1253.4\\[5pt]
C$_{110}$H$_{26}$ &        8.776&     108.1& 8.721&      49.5\\
&     8.420&     960.5& 8.389&      55.4& 8.497&    1155.5\\
&     7.796&    1774.1& 7.706&     124.1& 7.867&    1681.3\\
&     7.368&     345.2& 7.354&      27.2& 7.335&     350.0\\
&     6.983&     350.9& 7.029&      73.1& 7.016&     157.1\\
&     6.724&     500.6& 6.799&      13.8& 6.751&     118.1\\
&     6.479&     869.9& 6.590&      62.3\\
&     6.326&     738.8& 6.237&     175.6& 6.326&    1293.6\\[5pt]
C$_{112}$H$_{26}$ &       8.431&     644.0& 8.639&      66.6& 8.501&     516.5\\
& & & 8.168&      49.2& 8.190&     297.6\\
&     7.732&    1969.1& 7.882&      22.2& 7.795&    2046.0\\
&   & &  7.662&      23.5\\
&   & &  7.479&      26.4\\
&     7.241&     203.3& 7.254&      33.8& 7.250&      73.7\\
&     6.949&     197.1& 6.952&      33.5& 7.025&     132.6\\
&     6.745&     492.6& 6.730&      52.2& 6.775&     143.7\\
&     6.498&     483.8\\
&     6.308&     757.7& 6.285&     184.7& 6.339&     503.0\\[5pt]

C$_{130}$H$_{28}$ &       8.841&      90.3& 8.852&      59.8& 8.843&      74.6\\
&     8.472&    1593.4& 8.489&      27.9& 8.488&    1945.1\\
&     & & 8.193&      65.4\\
&     7.822&    2666.7& 7.899&      18.1& 7.862&    2515.4\\
&     & & 7.768&      20.1\\
&     7.245&     835.0& 7.231&      95.8& 7.133&     352.2\\
&     & & 6.975&      33.4\\
&     6.694&    1392.5& 6.660&      81.9& 6.713&     456.8\\
&     6.443&    1531.6\\
&     6.297&     811.5& 6.303&     209.1& 6.423&    1518.8\\[5pt]
\enddata
\end{deluxetable}

\clearpage

The peak wavelengths and integrated band strengths of the important
features in the 6-9 \mum\, region of the spectra for the neutral,
cation, and anion forms of the PAHs considered here are summarized in
Table~\ref{t2} and the spectra are shown in Fig.~\ref{fig_62}.  The
bands in the 6-9 \mum\, region of the spectra correspond to CC
stretching and CH in-plane bending vibrations.  When considering these
results, note that the intensity scale for the spectra of these very
large PAHs in the neutral form is at least 10 times smaller than those
of the cations and anions.  This behavior is consistent with that
found for all PAHs studied to date, namely the bands in the 6-9 \mum\,
region undergo significant intensity enhancement upon PAH ionization
(see Table~\ref{t2}).  Since the 6-9 \mum\, bands are rather weak for
the neutrals, this discussion will focus on the cations and anions.

Consider first the pure CC stretching region ($\sim$6.2 to 6.5 \mum)
in the spectra shown in Fig.~\ref{fig_62}.  The spectrum produced by
CC stretching vibrations (CC$_{str}$) of the cations is slightly richer
than that of the anions.  Nonetheless, the strongest band in this
region falls between 6.267 and 6.443 \mum\, in most cases (this
excludes \stf\, which has two nearly equal peaks at 6.326 and 6.479
\mum). The peak position of the CC$_{str}$ features in the spectra of the
anions also show less variation than the peak wavelength in the
spectra of the cations and they are slightly weaker in both an
absolute sense and relative to the bands between 7 and 9 \mum.  The
anion band peak tends to fall at somewhat longer wavelengths than that
of the cation, ranging from 6.326 to 6.423 \mum.
For the three largest PAH cations, there is a second peak in the range
6.443 and at 6.498 \mum\, and a third peak around $\sim$ 6.7 \mum. For
\sth$^+$, the 6.443 \mum\, component is stronger than the 6.297 \mum\,
component.  There is some surprising variation in the intensity of the
cation bands from one molecule to the next, and in this regard note
that this band is rather weak for \std$^+$.  There are a lot of modes
in this spectral range and those with significant intensity derive
that intensity from details of the molecular shape and the charge
distributed throughout the molecule.

Consider next the bands produced by vibrations involving PAH CC
stretching and CH in-plane bending vibrations.  These fall from about
7 to 9 \mum\, in the spectra shown in Fig.~\ref{fig_62}.  The features
in the 7 to 8 \mum\, region are associated with vibrations in which CC
stretching and CH in-plane bending motions are coupled, while those
between 8 and 9 \mum\, primarily arise from CH in-plane bending
vibrations.  The 7 to 9 \mum\, spectra of the cations and anions
resemble one another and, at first glance, the spectra of the three
largest PAHs appear similar.  However, upon closer inspection
important differences become evident. The following trends emerge when
comparing the spectra.

For the three largest PAHs, the strongest feature near 7.7 \mum\, (the
position of the strongest astronomcial feature) falls between 7.732
and 7.822 \mum\, for the cations and between 7.795 and 7.867 \mum\,
for the anions.  This tendency for the cation bands to fall at
slightly shorter wavelengths than for the anions holds for the smaller
PAHs considered here as well, although the spread is somewhat larger.

The variation in peak position of the strongest band near 7.7 \mum\,
is less pronounced for the anions than for the cations, mostly because
the bands in the small anions tend to fall at longer wavelengths than
those in small cations.  For example, for \std$^+$ the band falls
close to 7.67 \mum\,, but for \ste$^+$ the maximum is closer to 7.56
\mum.  For the corresponding anions, the difference is less than 0.05
\mum.  The 7.7 \mum\, band of the two smallest PAH cations, \stb$^+$,
and \stc$^+$ is not very pronounced.  However, further increasing the
size leads to a shift toward $\sim$7.8 \mum\, for the three largest
PAH cations considered here.

We turn now to the bands that arise primarily from CH in-plane bending
vibrations, the features in the 8 to 9 \mum\, region of the spectra in
Fig.~\ref{fig_62}.  Overall, the spectral features of both the anions
and cations in this region are rather similar.  While there is slight
variation in the position of these bands, there appears a general
shift to longer wavelength with increasing size of the PAH cation.
There is a band in the 8.279 to 8.467 \mum\, range for the three
smallest PAHs, but it is not very strong.  However, as PAH size
increases, the intensity of this band increases significantly and it
shifts to longer wavelength, producing the very prominent band that
appears near 8.5 \mum\, in the spectra of the 4 largest PAHs for both
the cations and anions.

For the molecules studied in this work, there are many allowed modes
throughout the spectra.  It is therefore interesting to note that for
some molecules, a few bands carry all the intensity in the 6-9 $\mu$m
region, while for others there are several bands with significant
intensity.  This is very different from the C-H stretching region
where the band position varies little for the molecules studied here.
Thus, the 6-9 \mum\, region of the spectra may yield some insight into
the character of the molecules present in the astrophysical
environment. With this in mind, we have viewed the modes that compose
the bands that carry significant intensity, for both the molecules in
this work and the lower symmetry species presented in paper II.  While
at the present time we are unable to make a definitive prediction of
the nature of the spectra in the 6-9 \mum\, region based on molecular
shape, a study of the modes suggests that the differences in the
spectra are related to the shape of the edge.  We continue to
investigate this and will revisit this in paper II.

\subsection{The CH Out-of-Plane Bending Vibrations (9 to 15 \mum).}
\label{lab_112}

\clearpage

\begin{deluxetable}{l@{\hspace{15pt}}rr@{\hspace{15pt}}rr@{\hspace{15pt}}rr}
\tablecaption{\label{t3} The 9-15 $\mu$m band position maxima ($\lambda$,
in $\mu$m) and total intensity (I, in km/mol).  For PAHs with {\it
any} band stronger than 10km/mol, all charge states are listed. If the
bands for the cation, neutral and anion state of a PAH all fall below
10km/mol they are not included. }
\tablehead{Molecule & \multicolumn{2}{l}{Cation} &\multicolumn{2}{l}{Neutral} & \multicolumn{2}{l}{Anion} \\
 &  \multicolumn{1}{c}{$\lambda$} &  \multicolumn{1}{c}{I} &  \multicolumn{1}{c}{$\lambda$} &  \multicolumn{1}{c}{I}&   \multicolumn{1}{c}{$\lambda$} &  \multicolumn{1}{c}{I}}
\startdata
C$_{24}$H$_{12}$ &       13.098&      13.7& 12.910&      11.7& 13.158&      16.5\\
&    12.757&      32.7& & & 12.835&      17.5\\
&    12.422&       7.0& & & 12.323&     164.8\\
&    11.355&     190.5& 11.575&     175.6& 10.752&       2.2\\[3pt]

 C$_{54}$H$_{18}$  &      13.922&      21.9& 13.864&      21.7& 13.922&      17.7\\
&    12.668&      53.9& 12.788&      43.3& 13.144&      23.3\\
&    11.939&      13.2& 11.916&      10.2& 11.710&     107.9\\
&    10.881&     218.1& 11.061&     221.1& 11.353&     128.9\\
&    10.162&      46.6& 10.167&       2.8& 10.195&       1.4\\
&     9.768&       1.3&   9.790&      12.6& 9.835&       8.1\\
&     9.186&      11.3\\[3pt]

C$_{66}$H$_{20}$ &        12.868&       1.7& 12.702&      52.7& 12.893&      35.1\\
&    12.596&      57.4& & & 12.466&       1.3\\
&    12.136&      10.9& 12.270&       2.6& 12.148&       3.4\\
&    11.307&      17.4& 11.682&       7.5& 11.567&      35.0\\
&    10.984&      48.1& 11.150&      39.1& 11.355&     232.7\\
&    10.823&     194.7& 10.995&     216.3& 10.989&       6.0\\[3pt]

C$_{78}$H$_{22}$ &  \  14.102&      13.5& 14.112&      10.4& 14.178&       7.5\\
&    12.653&      53.1& 12.780&      44.4& 12.974&      27.7\\
&    11.955&       5.4& 11.942&      18.5& 11.969&       6.8\\
&    11.722&      21.9& 11.570&       4.7& 11.783&       1.8\\
&    10.971&      76.5& 11.132&      70.5& 11.302&     301.4\\
&    10.810&     198.5& 10.966&     212.2& 10.728&       8.0\\
&    10.206&      13.6& 10.241&       5.7& 10.262&       6.9\\
&     9.347&       9.0& 9.387&       2.0& 9.387&      16.0\\[3pt]

C$_{96}$H$_{24}$ &     14.648&       9.5& 14.639&       5.7& 14.786&      13.8\\
&    13.203&       8.6& 13.158&       6.0& 13.293&      10.3\\
&    12.677&      17.4\\
&    12.385&      76.9& 12.477&      74.2& 12.606&      65.8\\
&  & & & & 11.338&     162.7\\
&    10.825&     299.0& 10.959&     316.4& 11.120&     199.6\\
&     9.630&      29.7& 9.630&      21.4\\
&     9.334&      49.3& 9.341&       3.2& 9.369&      18.9\\[3pt]

 C$_{110}$H$_{26}$ &       14.535&      16.2& 14.539&      13.1& 14.560&      11.5\\
&    13.040&       9.3& 13.098&       6.2& 12.933&      70.5\\
&    12.655&      55.4& 12.749&      46.1& 12.557&       6.8\\
&    12.090&       4.2& 12.070&       4.3& 12.053&      51.6\\
&    11.686&      28.0& 11.820&      25.6& 11.741&       3.7\\
&    11.400&       8.0& 11.370&       7.8& 11.549&      20.1\\
&    11.183&      14.7& & & 11.177&     350.4\\
&    11.023&      26.1\\
&    10.814&     316.7& 10.954&     348.6& 10.875&       3.7\\
&    10.134&      11.0& 10.128&       4.8& 10.148&       3.1\\
&     9.458&      45.4& 9.349&      22.4& 9.461&      49.8\\[3pt]

C$_{112}$H$_{26}$ &    12.466&      67.0& 12.544&      54.1& 12.692&      53.3\\
&    12.099&       7.3& 12.109&       4.4& 12.226&      17.8\\
&    11.716&      15.2& 11.846&      12.9& 11.640&      22.3\\
&    10.820&     324.3& 10.941&     341.2&11.148&     356.1\\
&    10.122&      15.7& 10.139&       4.8& 10.301&      16.1\\
&     9.620&      16.1& 9.620&      17.5& 9.547&      20.9\\
&     9.291&      38.8& 9.297&       8.1& 9.322&      14.1\\[3pt]

C$_{130}$H$_{28}$ &    14.019&       9.1& 14.071&       7.4& 14.233&      29.3\\
&    12.587&      66.5& 12.703&      59.6& 12.706&      69.5\\
&    & & & & 12.047&      16.4\\
&    & & & &11.902&      44.4\\
&    11.660&      36.1& 11.802&      29.1& 11.610&      25.9\\
&    11.085&      10.2& 11.498&       9.0& 11.116&     383.5\\
&    10.817&     351.2& 10.936&     369.8\\
&    10.138&      21.9& 10.167&       8.8& 10.166&      54.1\\
&     9.836&      26.0& & & 9.853&       7.6\\
&     9.575&      23.7& 9.574&      12.3& 9.602&       5.0\\[3pt]
\enddata
\end{deluxetable}

\clearpage

\begin{table}
  \caption{\label{t4} The total intensity (I, in km/mol) and intensity
  per CH (I(CH), in km/mol) for the solo and duo modes.}
\begin{center}
\begin{tabular}{l@{\hspace{7pt}}l@{\hspace{7pt}}rr@{\hspace{15pt}}rr@{\hspace{15pt}}rr}
\hline\\[-5pt]
Molecule & Mode & \multicolumn{2}{c}{Cation} &\multicolumn{2}{c}{Neutral} & \multicolumn{2}{c}{Anion} \\
& & \multicolumn{1}{c}{I} & \multicolumn{1}{l}{I(CH)}& \multicolumn{1}{c}{I} & \multicolumn{1}{l}{I(CH)}& \multicolumn{1}{c}{I} & \multicolumn{1}{l}{I(CH)}\\[5pt]
coronene 
&   solo & 220.67 & 36.78 &  221.08 &  36.85 & 128.89 &  21.48 \\
&   duo1 &  14.89 &  1.24 &   10.28 &   0.86 & 114.67 &   9.56 \\
&   duo2 &  53.86 &  4.49 &   43.33 &   3.61 &  23.33 &   1.94 \\[5pt]
\stc
&   solo & 260.12 & 32.51 &  255.46 &  31.93 & 238.30 &  29.79 \\
&   duo1 &  69.11 &  5.76 &   10.08 &   0.84 &  40.22 &   3.35 \\
&   duo2 &   1.74 &  0.14 &   52.71 &   4.39 &  35.05 &   2.92 \\[5pt]
\std
&   solo & 275.02 & 27.50 &  282.71 &  28.27 & 301.39 &  30.14 \\
&   duo1 &  34.34 &  2.86 &   25.41 &   2.12 &  12.20 &   1.02 \\
&   duo2 &  59.44 &  4.95 &   50.50 &   4.21 &  30.84 &   2.57 \\[5pt]
\ste
&   solo & 299.05 & 24.92 &  316.42 &  26.37 & 362.34 &  30.20 \\
&   duo1 &  82.03 &  6.84 &   76.98 &   6.42 &   1.56 &   0.13 \\
&   duo2 &  17.53 &  1.46 &    6.01 &   0.50 &  65.78 &   5.48 \\[5pt]
\stf
&   solo & 365.42 & 26.10 &  356.38 &  25.46 & 354.08 &  25.29 \\
&   duo1 &  32.27 &  2.69 &   29.85 &   2.49 &  82.29 &   6.86 \\
&   duo2 &  64.72 &  5.39 &   52.29 &   4.36 &  70.56 &   5.88 \\[5pt]
\stg
&   solo & 325.01 & 23.22 &  341.21 &  24.37 & 356.13 &  25.44 \\
&   duo1 &  89.47 &  7.46 &   79.95 &   6.66 &  40.12 &   3.34 \\
&   duo2 &   7.85 &  0.65 &   15.81 &   1.32 &  61.65 &   5.14 \\[5pt]
\sth
&   solo & 361.41 & 22.59 &  369.77 &  23.11 & 383.45 &  23.97 \\
&   duo1 & 102.59 &  8.55 &   38.10 &   3.17 &  86.72 &   7.23 \\
&   duo2 &   6.36 &  0.53 &   66.25 &   5.52 &  77.49 &   6.46 \\[5pt]
\hline
\end{tabular}
\end{center}
\noindent
\end{table}

\clearpage

\begin{figure*}
   \begin{minipage}[c]{0.33\textwidth}
      \centering \includegraphics[width=\textwidth]{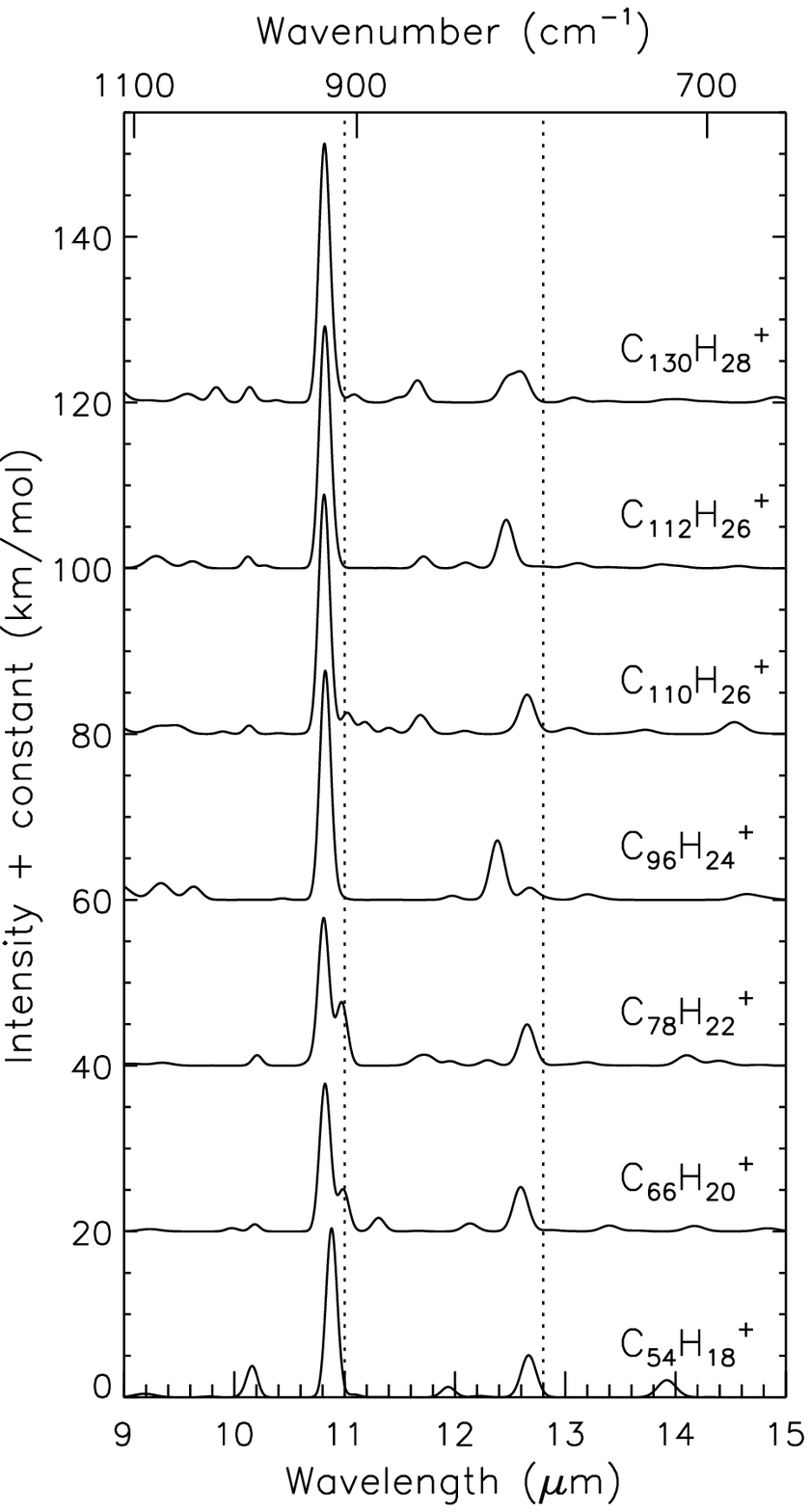}
   \end{minipage}
   \begin{minipage}[c]{0.33\textwidth}
      \centering \includegraphics[width=\textwidth]{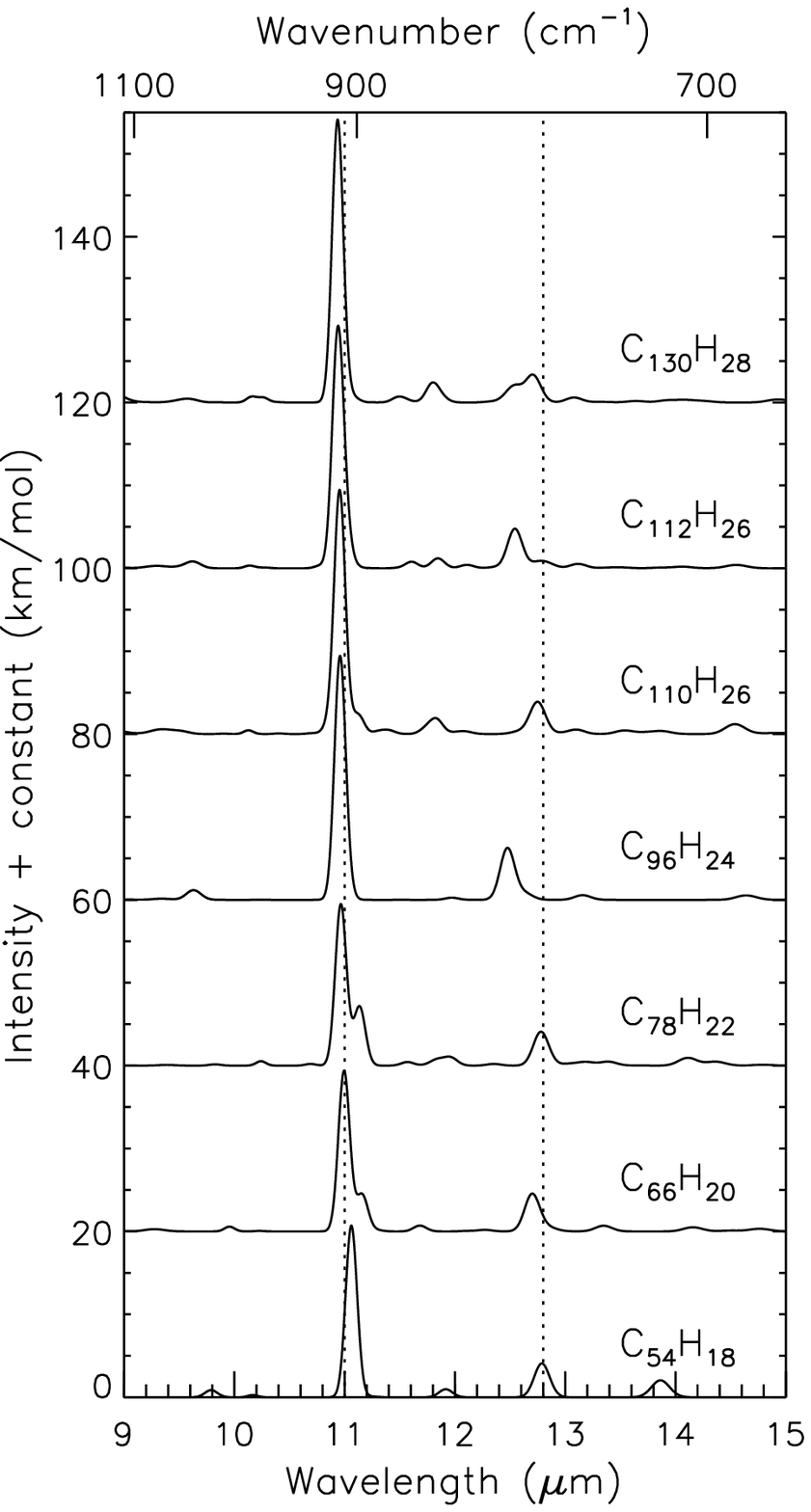}
   \end{minipage}
   \begin{minipage}[c]{0.33\textwidth}
      \centering \includegraphics[width=\textwidth]{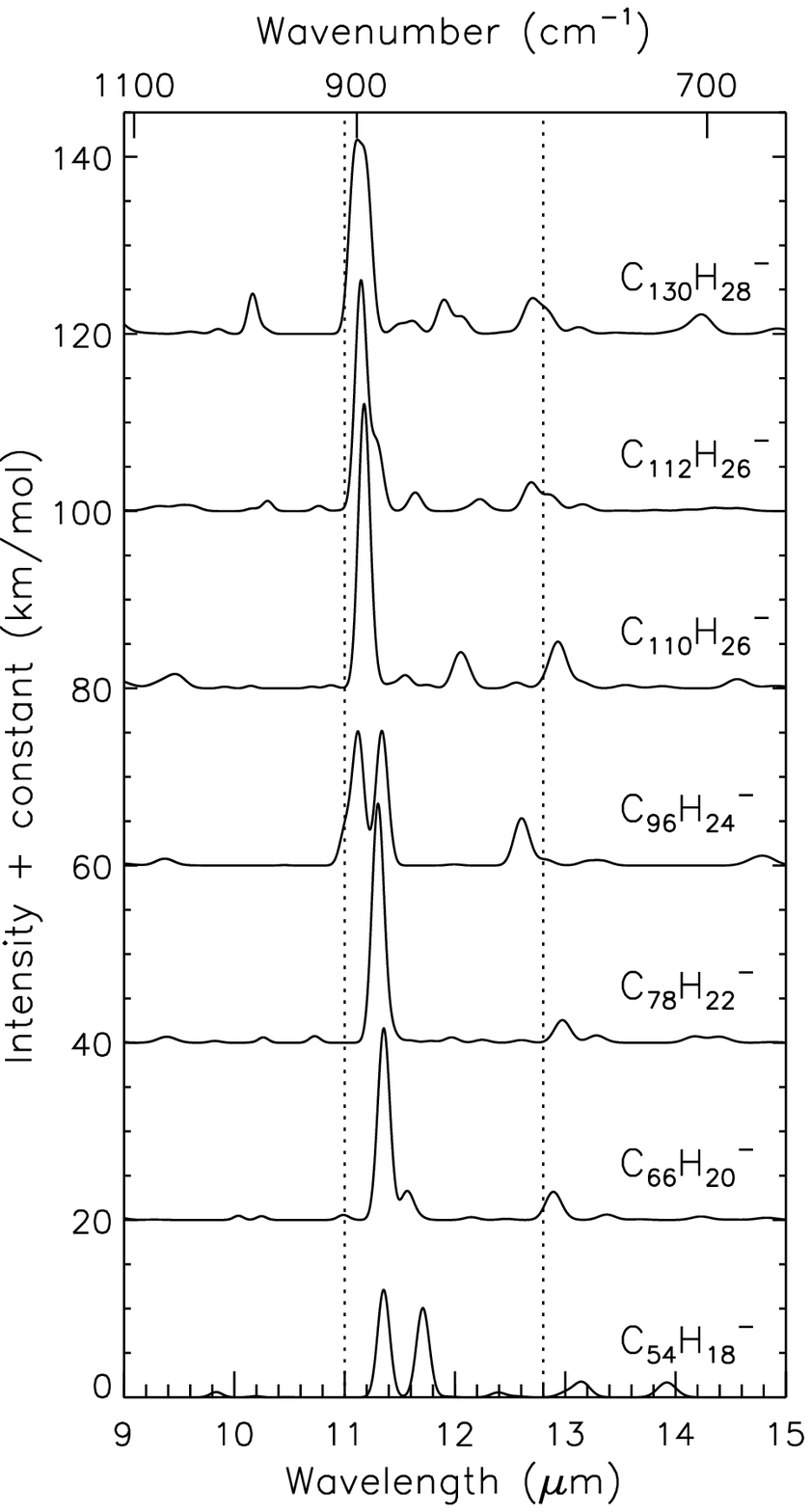}
   \end{minipage}
\caption{The synthetic absorption spectra in the 9 to 15 $\mu$m\,
region for the large symmetric PAH cations, neutrals and anions
considered here. To guide the eye, dotted lines at 11.0 and 12.8
$\mu$m\, are also shown.}
\label{fig_112}
\end{figure*}

The 9 to 15 \mum\, region of the spectra for the neutral, cation, and
anion forms of the PAHs considered here are shown in
Fig.~\ref{fig_112} and the peak wavelengths and integrated band
strengths of the most significant bands in these spectra are
summarized in Table~\ref{t3}.  The bands shown in Fig.~\ref{fig_112}
correspond to CH out-of-plane bending vibrations (CH$_{oop}$).  In contrast
with the spectra in the 5 to 9 \mum\, region, the intensities of the
CH$_{oop}$ bands for all three PAH charge forms are very similar.  In all
three cases, the larger the PAH the shorter the wavelength of the
strongest feature.

Perusal of Fig.~\ref{structure} shows that there are only solo and duo
hydrogens on the PAHs considered here.  For these PAHs, the CH$_{oop}$
band for the solo hydrogens falls between 10.936 and 11.061 \mum\, for
the neutral forms, between 10.810 and 10.881 \mum\, for the cations,
and between 11.116 and 11.353 \mum\, for the anions (see
Fig.~\ref{fig_112}). These all fall within the broader range for solo
hydrogens on smaller PAHs \citep[][and references
therein]{Hudgins:tracesionezedpahs:99}.  However, this generalization
doesn't hold for the bands produced by duo hydrogen vibrations on
these larger PAHs.  Analysis of the vibrations of these larger PAHs
shows that the CH$_{oop}$ vibrations for duo hydrogens produces two
bands: a weaker one between $\sim$11.5 and 12 \mum\, and a stronger
one near $\sim$12.8 \mum.  The weaker band lies in the region normally
associated with duo CH$_{oop}$ vibrations while the more prominent
band falls in the region normally attributed to trio adjacent
hydrogens, yet none of these PAHs have trio hydrogens.  Furthermore,
the weak duo band peaking near 11.5-12 \mum\, shows less variation in
peak position than does the duo band near $\sim$12.8 \mum.  The
strongest of the bands produced by duo CH$_{oop}$ bends for the
neutral and cation species considered here fall just below 12.8 \mum\,
while half of the PAH anion duo band peaks lie above 12.8 \mum\, and
half below.  Interestingly, the spectrum of coronene (\sta), a small
PAH with only duo hydrogens, shows a prominent band at 11.575 \mum\,
as well as a weak band at 12.910 \mum\, \citep[see
Table~\ref{t3}][]{Hudgins:tracesionezedpahs:99}.

The intensities of these bands are listed in Table \ref{t4}.
Inspection of this Table \ref{t4} reveals trends in the A-values per
CH for the bands produced by the CH$_{oop}$ vibrations that differ
significantly from those in small PAHs.  Excluding the entries for
coronene, the average A value per CH for the solo
hydrogens\footnote{10.8-11.4 \mum\, range} in the cation, neutral, and
anion charge states of the very large PAHs are 26 km/mol, 27 km/mol,
and 27 km/mol respectively.  These values are about twice the average
A values for the neutral (13 km/mol) and cation forms (14 km/mol) of
smaller PAHs \citet{Hony:oops:01}.  Similarly, the average for the duo
modes\footnote{11.4-12.6 and 12.6-13.2 \mum\, ranges} of the very
large PAHs considered here are 7.7 km/mol, 7.0 km/mol, and 7.2 km/mol
for the cation, neutral and anion forms versus 2.5 km/mol and 2.4
km/mol for the cation and neutral forms of the smaller PAHs.  The
ratio of solo-to-duo hydrogens in the PAHs considered here spans the
range from 1.3 (\sth) to 0.5 (\stb).  The relative intensities of the
corresponding bands in Figure \ref{fig_112} reflects the constancy of
the A-values; independent of the PAHs in this sample.  It is important
to note that the A-values for the duo modes in the smaller PAHs were
determined from PAHs with structures that included solo, trio, and
quadrupule hydrogens.  The larger, symmetric, compact PAHs discussed
here are quite different in this respect.  As they are of a size
comparable to those that dominate the astronomical PAH mix, the
conclusions derived here and in paper II are more appropriate for the
emitting astronomical PAH population.

\begin{figure}
\plotone{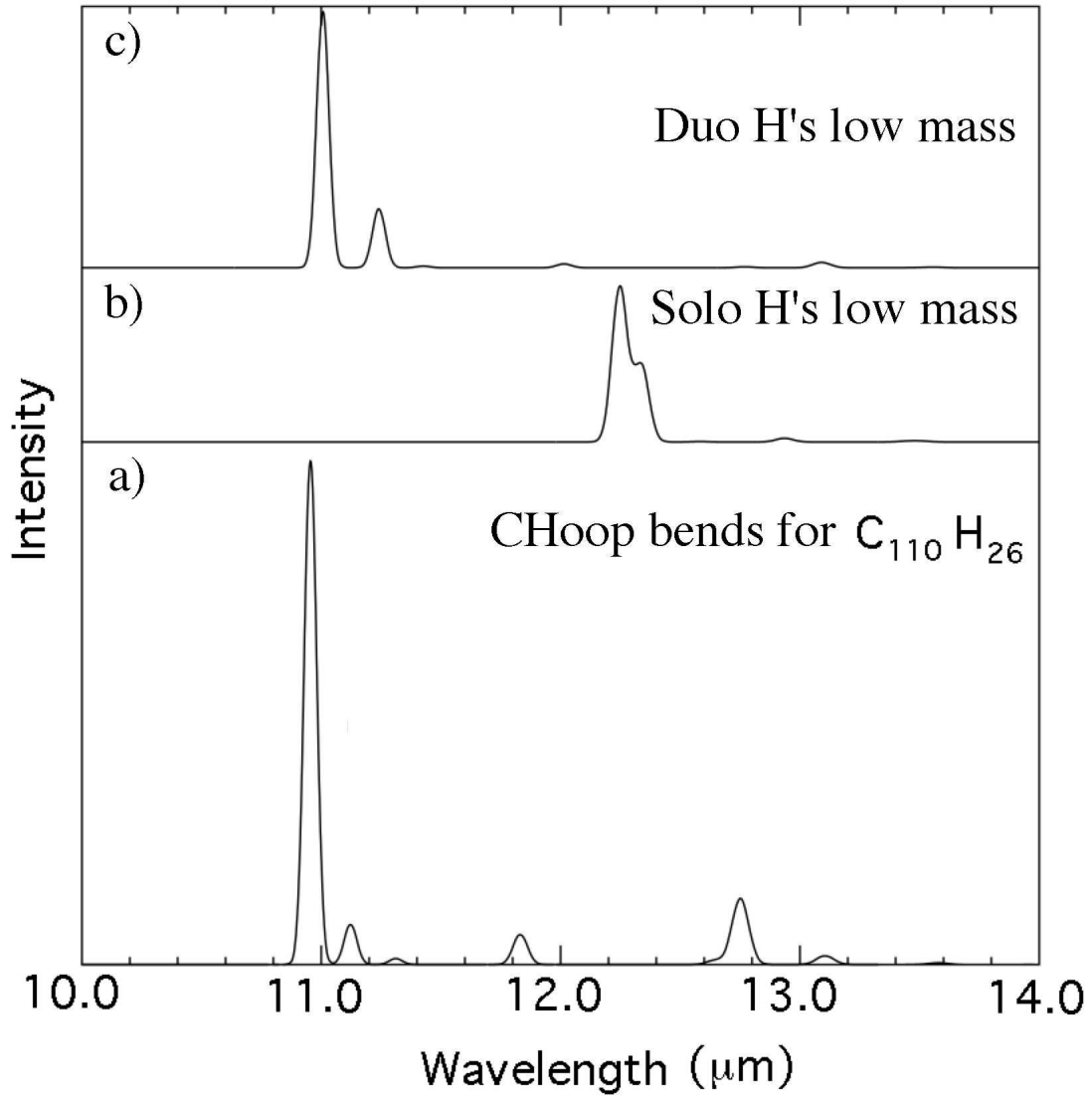}
\caption{The influence of hydrogen mass on the CH$_{oop}$ bands
of C$_{110}$H$_{28}$. a) The spectrum of normal C$_{110}$H$_{28}$; b)
The computed spectrum with the mass of the solo hydrognes artificially
reduced to 0.2 AMU; c) The computed spectrum with the mass of the duo
hydrogens artificially reduced to 0.2 AMU.}
\label{fig_oops_D}
\end{figure}

To gain more insight into the origin of the variation of the duo
hydrogen modes, the force constants for neutral \stf\, were used to
recompute the vibrational spectra, but with the mass of either the solo
or duo hydrogens changed from 1.00783 (hydrogen) to 0.2 AMU.  While a
mass of 0.2 AMU is completely artificial, this change in mass shifts
the CH$_{oop}$ modes of the altered mass hydrogens out of the 9-15 \mum\,
region and eliminates the vibrational coupling between the solo and
duo hydrogens.  The result of this numerical experiment is shown in
Fig.~\ref{fig_oops_D}. Note that the FWHM is reduced to 7 cm$^{-1}$ in
these spectra to more clearly show the changes.  Comparing the
spectrum of \stf\, with that in which the mass of the duo hydrogens is
reduced shows that the coupling with the duo hydrogens increases the
intensity of the solo peak at 11 \mum\, and shifts it very slightly in
wavelength.  The small solo peak near 11.2 \mum\, is reduced in
intensity and shifted slightly to shorter wavelength when coupled to
the duo hydrogens.  The duo bands without coupling are reasonably
strong and fall near 12.3 \mum, but when coupled to the solo band,
they lose intensity and are split into two bands, one near 11.8 \mum\,
and the other at 12.8 \mum. The duo bands show a much more dramatic
change with coupling than the solo bands for these large compact PAHs
which have comparable amounts of solo and duo hydrogens.

\section{Astrophysical Implications }
\label{astro}

The mid-IR spectroscopic properties of the large PAHs presented above
will now be compared with observations.  The discussion parallels the order
in Sect.~\ref{lab}.

\begin{figure*}
   \begin{minipage}[c]{0.33\textwidth}
      \centering \includegraphics[width=\textwidth]{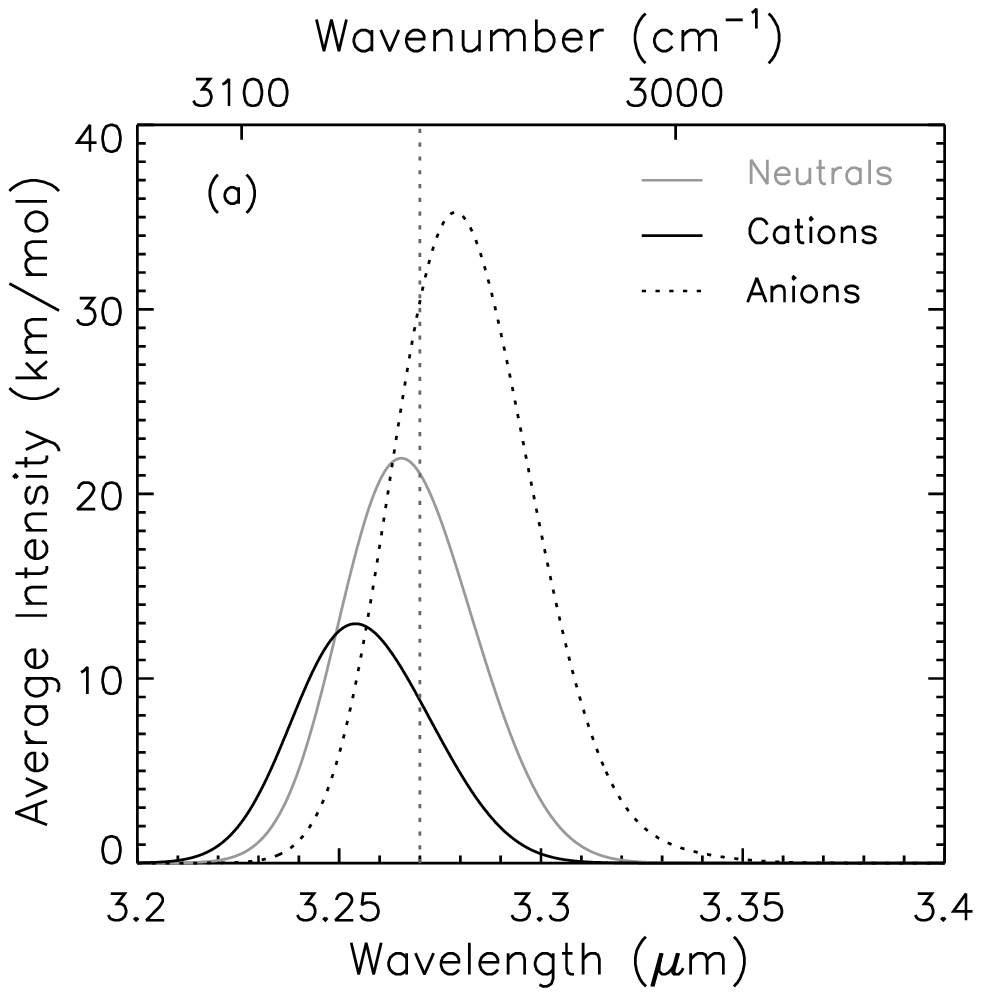}
   \end{minipage}
   \begin{minipage}[c]{0.33\textwidth}
      \centering \includegraphics[width=\textwidth]{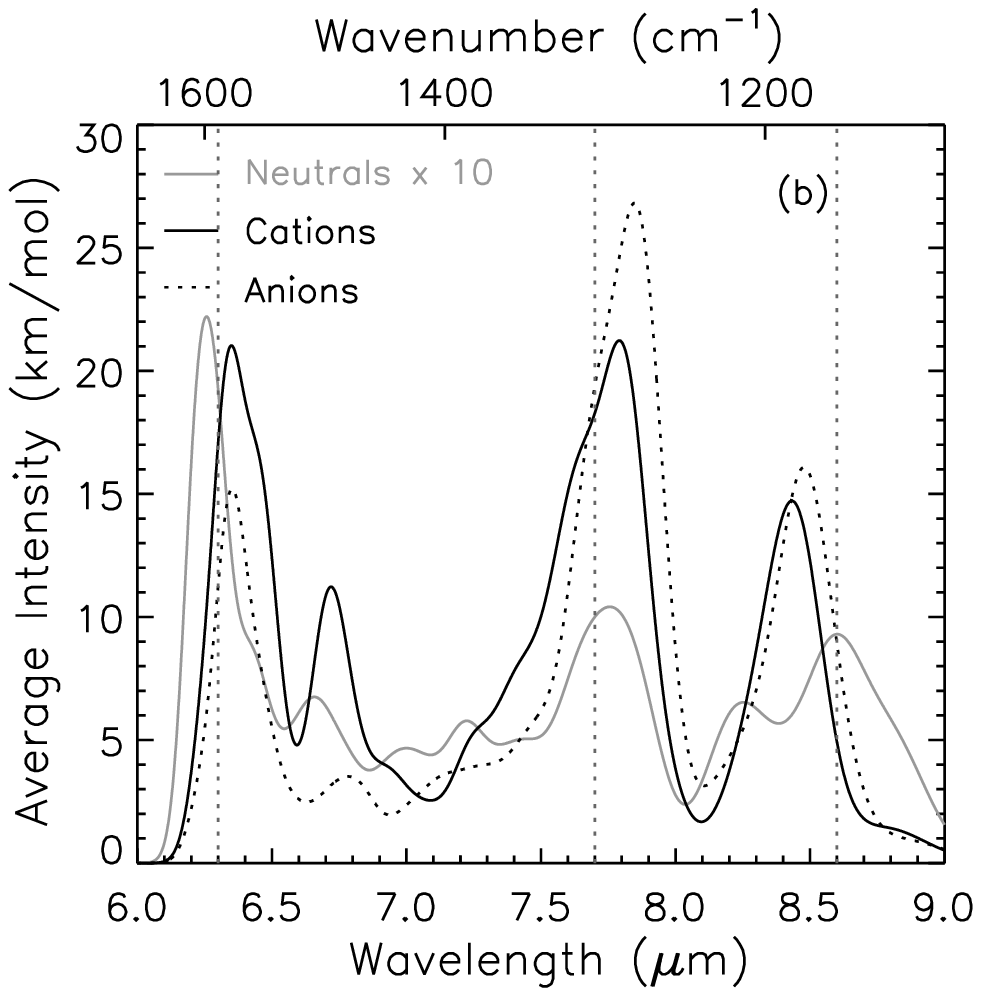}
   \end{minipage}
   \begin{minipage}[c]{0.33\textwidth}
      \centering \includegraphics[width=\textwidth]{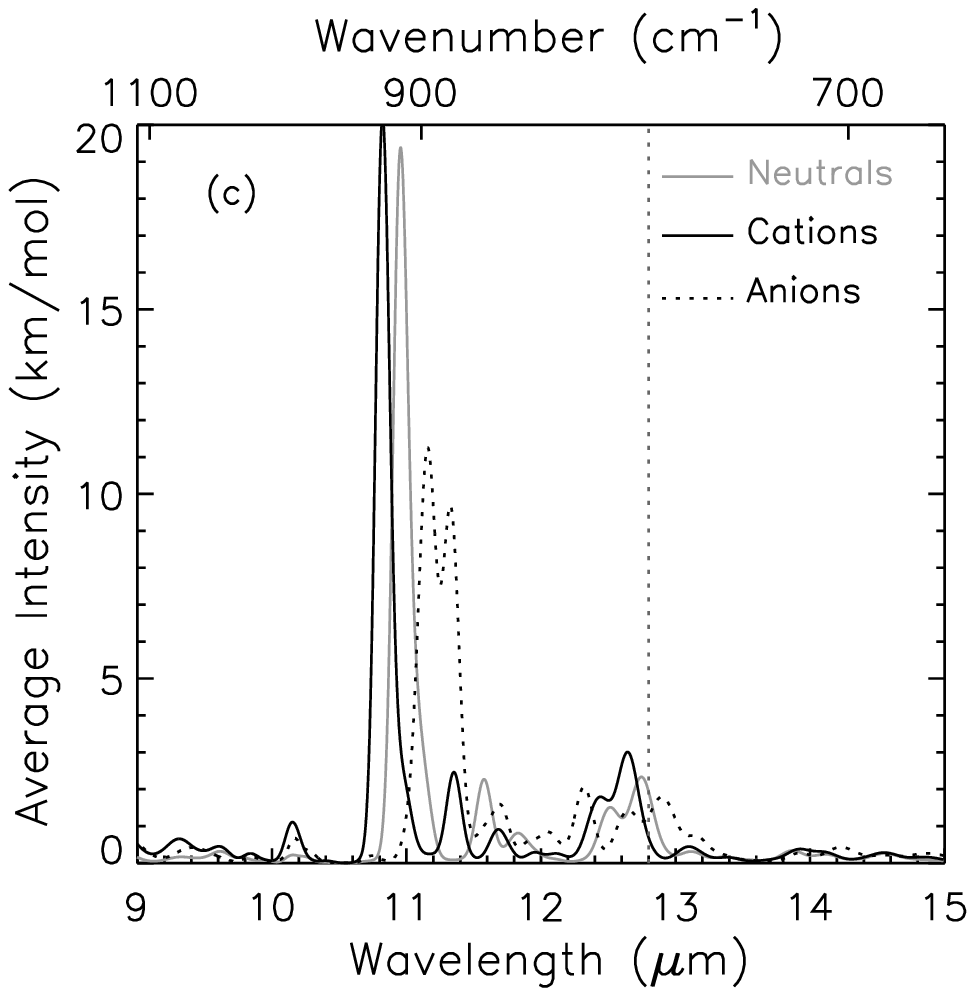}
   \end{minipage}
\caption{The average of the synthetic absorption spectra for the
large symmetric PAH cations, neutrals and anions considered here. To guide
the eye,  dotted lines at 3.27, 6.3, 7.7, 8.6 and 12.8 $\mu$m\, are
also shown.}
\label{fig_average}
\end{figure*}

\begin{figure}
      \centering \includegraphics[width=.45\textwidth]{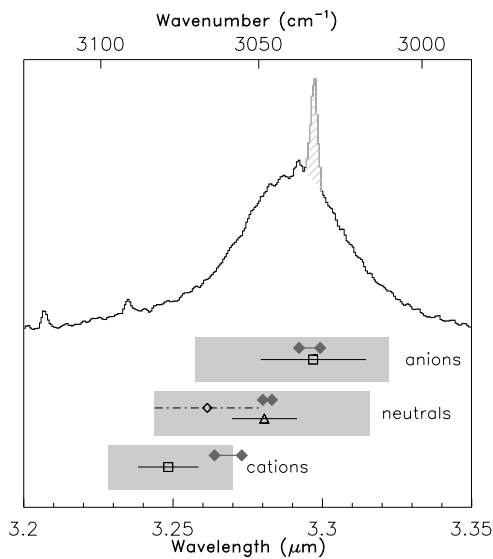}
\caption{Comparison of the astronomical emission feature from NGC~7027
(class A$_{3.3}$) with the peak positions of the CH stretch in various
PAHs.  The striped, narrow grey feature in the spectrum of NGC7027 is
the Pf$\delta$ line. The PAH data are redshifted by 15 cm$^{-1}$, a
shift which is intrinsic to the emission process (see text for
details). The connected, filled diamonds indicate the range of the CH
stretch in the large PAHs considered here. The shaded horizontal bars
indicate the full range of the PAH CH stretching modes for the smaller
PAH sample described in the text (Sect.~\ref{astro_33}). The
symbols crossed by short horizontal lines represent the average
wavelength and standard deviation for that bar. The open diamond and
triangle in the bar for neutral PAHs show the average and standard
deviations for peak positions of PAHs containing $\ge$ 40 and $<$ 40 C
atoms, respectively, in the previous data set. Figure adapted from
\citet{vanDiedenhoven:chvscc:03}.}.
\label{fig_chstr}
\end{figure}
\subsection{The C-H stretching vibrations (2.5 - 3.5 \mum).}
\label{astro_33}

The astronomical emission bands arising from the CH stretching
vibrations were discussed in detail by
\citet{vanDiedenhoven:chvscc:03}.  These authors compared their
astronomical data with experimental and theoretical spectra of over
100 small PAHs ($<$ C$_{40}$) in their neutral and positively charged
states, supplemented by the spectra of a few larger PAHs, and 27 small
PAH anions.  This analysis is extended here to include the impact of
the spectrum of the larger PAHs shown in Fig.~\ref{structure} on the
earlier conclusions.

Fig.~\ref{fig_average}a shows the feature produced by averaging the
bands associated with each charge state previously presented in
Fig.~\ref{fig_33}.  This clearly demonstrates the shift in peak
position as a function of PAH charge.  Fig.~\ref{fig_chstr} shows the
3.29 \mum\, astronomical emission feature from NGC 7027 along with
grey bars which represent the full range of PAH CH stretching peak
positions for the neutral, cationic, and anionic PAHs in the sample
described above.  Superposed on this figure is the range in peak
positions for the very large compact PAHs presented in Table~\ref{t1}
(dark grey bar connecting filled diamonds).  To facilitate the
comparison between the spectra presented here and the astronomical
spectra, all laboratory and theoretical spectra have been redshifted
by 15 cm$^{-1}$ to take the redshift into account that is intrinsic to
the emission process \citep{Flickinger:91, Brenner:benz+naph:92,
Colangeli:T:92, Joblin:T:95, WilliamsLeone:95, Cook:excitedpahs:98}.
The peak positions for compact, very large PAHs fall within the
regions spanned by the earlier sample, but span a much narrower range.
Consequently, the principle conclusion presented in
\citet{vanDiedenhoven:chvscc:03} for this feature, namely that its
peak position is most consistent with an origin in neutral and
negatively charged PAHs, while clearly ruling out an origin in PAH
cations, remains unaltered. The proximity of the ranges in the
CH$_{str}$ peak position for the neutral and anion PAH forms suggests
that the two classes of astronomical 3.3 \mum\, band designated as
A$_{3.3}$ and B$_{3.3}$ by \citet{vanDiedenhoven:chvscc:03} reflect
the relative contributions of PAH neutrals and anions. 

The narrow range in CH stretch frequencies listed in Table~\ref{t1}
and illustrated in Fig.~\ref{fig_chstr}, and the predominance of one
feature in the CH stretching region are characteristics inherent in
the spectroscopy of compact and largely symmetric PAHs.  Such
structures require that the molecules have similar edge structures.
This, in turn, leads to similar spectra involving the CH modes as the
local environment of each peripheral CH group is similar along the
edge of the molecule within a given molecule and from one molecule to
the next.  The similarity between the absorption band of the CH
stretch in the large compact PAHs described here and the astronomical
3.29 \mum\, feature suggests that compact and somewhat symmetric PAH
cations and anions dominate the PAH mixture producing this
astronomical feature. This aspect of very large PAH spectroscopy will
be further discussed in paper II.

\subsection{The C-C Stretching and C-H in-plane bending vibrations (5 - 9 \mum).}
\label{astro_62}

The 5 to 9 \mum\, astronomical emission bands arising from PAH CC
stretching and CH in-plane bending vibrations were discussed in great
detail by \citet{Peeters:prof6:02}.  Their analysis was based on a
dataset of experimental and theoretical spectra similar to that
described in Sect.~\ref{astro_33} above and is extended here to include
the spectroscopy of the larger PAHs shown in Fig.~\ref{structure}.
Fig.~\ref{fig_average}b shows the average 6 - 9 \mum\, spectra of
these PAHs in their neutral, cation, and anion forms.  The
corresponding spectra of the neutral PAHs will not be considered
further here as their absorption band intensities are at least an
order of magnitude smaller than those of the corresponding cations and
anions.

Consider first the spectroscopy near 6.2 to 6.3 \mum, the region
attributed to pure CC stretching vibrations.  Summarizing
Sect. \ref{lab_62}, the CC$_{str}$ features in the spectra of the anions
show less variation than the spectra of the cations and they are
slightly weaker in both an absolute sense and relative to the bands
between 7 and 9 \mum.  Nonetheless, Figure \ref{fig_average}b shows
that the average spectrum of the cations and anions is dominated by a
band that peaks near 6.3 \mum, close to the well-known astronomical
emission band that peaks between 6.2 and 6.3 \mum, depending on the
object type \citep{Peeters:prof6:02}.  The spectra presented here
extend our earlier finding, that the pure CC stretch in PAHs comprised
of only C and H cannot reproduce the observed peak position of
corresponding astronomical feature, to very large PAHs (note that the
redshift of 15 cm$^{-1}$ is not applied to Figure \ref{fig_average} and
hence in emission, the bands will fall longwards of the astronomical class A 6.2
PAH band).  This supports our previous conclusion that pure PAH
molecules cannot reproduce the 6.2 \mum\, component \citep[Class A
in][]{Peeters:prof6:02} of the astronomical emission feature, leading
to the suggestion that astronomical PAHs contain nitrogen
\citep[PANHs,][]{Peeters:prof6:02, Hudgins:05}.  In this regard, it is
important to recall that the 6 - 9 \mum\, bands in neutral PANHs have
absorption strengths half as large as do PAH ions.

It is interesting to note that the cations and anions of several of
these large symmetric species have bands in the 6.7-6.8 \mum\, region,
in particular for \ste\, and \sth.  Bands in this region are often
thought to arise from CH deformations in aliphatic side-groups. To
better understand the origin of the bands in the region, more
calculations on large PAHs are required.

Turning to the features near 7.7 \mum, Fig.~\ref{fig_62} shows that
the cations and anions of the PAHs considered here have strong bands
close to this position.  For the cations, as the PAH size increases
this band tends to shift to longer wavelength but there is some
variation in position with PAH.  For the anions, the band seems to
fall closer to 7.8 \mum\, and is less dependent on the size of the
PAH.  The average spectra in Figure \ref{fig_average}b show that the
strong feature produced by both the cations and anions overlap to a
large extent, but that the average anion peaks about 0.1 \mum\, to the
red of the average cation peak.  Together, these and earlier data
suggest that the astronomical 7.7 \mum\, band is produced by a mixture
of small and large PAH cations and anions, with small PAHs
contributing more to the 7.6 \mum\, component and large PAHs more to the 7.8
\mum\, component. \citet{Peeters:prof6:02} showed that when the 7.8
\mum\, component dominates the 7.7 \mum\, complex, its peak position
varies between 7.8 and 8 \mum. Given the slight difference in peak
position between large PAH cations and anions, this suggests that
negatively charged PAHs contribute more to the red portion of the 7.8
\mum\, component than do PAH cations. 

Lastly, for both anions and cations, a band near 8.5 \mum\, grows in
strength as PAH size increases (Figure \ref{fig_62}).  Figure
\ref{fig_average}b shows that the bands from both the large PAH anions
and cations overlap, again with the anions contributing more strongly
at slightly longer wavelengths.  \citet{Peeters:prof6:02} found a
correlation between the peak of the astronomical 7.8 \mum\, component
of the 7.7 \mum\, feature, and the peak position of the 8.6 \mum\,
band, i.e. they both are redshifted by a similar relative degree.  The
failure of small PAHs to reproduce i) the 7.8 \mum\, emission
component, ii) a prominent 8.6 \mum\, emission feature and iii) the
correlation between the peak position of the astronomical 7.7 \mum\,
feature and the 8.6 \mum\, band suggest that both the 7.8 \mum\,
component and the 8.6 \mum\, band originate primarily in large,
cationic and anionic PAHs, with the specific peak position and profile
reflecting the particular cation to anion concentration ratio in any
given object.

\begin{figure}
      \centering \includegraphics[width=.45\textwidth]{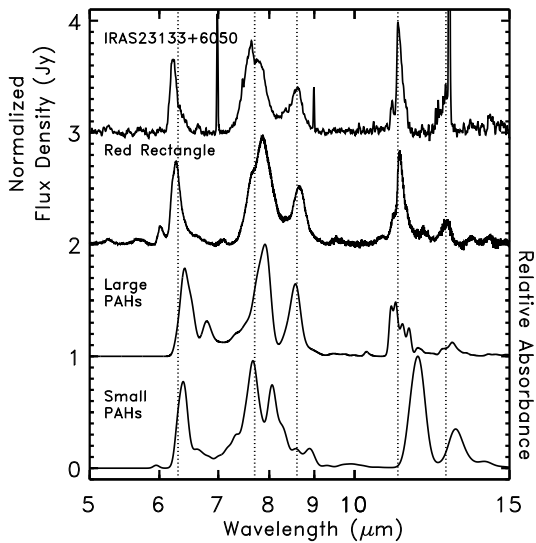}
\caption{Astronomical PAH emission spectra from 5 to 15 $\mu$m
compared with the spectra of PAH mixtures. Shown are the continuum
subtracted ISO-SWS spectrum of IRAS23133+6050 representing class A PAH
profiles {\bf (top)}; the continuum subtracted ISO-SWS spectrum of the
Red Rectangle, representing class B PAH profiles {\bf (second)}; the
average spectrum of the large symmetric PAH cations, neutrals and
anions considered here {\bf (third)}; and the composite absorption
spectrum generated by co-adding the individual spectra of 11 PAHs,
reproduced from \citet[][see text for composition]{Peeters:prof6:02}
{\bf (bottom)}.
A redshift of 15~cm$^{-1}$, intrinsic to the emission process, has
been applied to the positions of the large and small PAH spectra. To
guide the eye, dotted vertical lines are also shown.}
\label{fig_6to9}
\end{figure}

Fig.~\ref{fig_6to9} compares the emission spectrum from the HII region
IRAS~23133+6050 and the Red Rectangle to the average spectrum produced
by the large PAHs computed here and an earlier ``best fit'', average
mixture of smaller PAHs \footnote{The composite spectrum of 11 PAHs
consists out of 22\% neutral coronene (C$_{24}$H$_{12}$); 19\%
3,4;5,6;10,11;12,13-tetrabenzoperopyrene cation (C$_{36}$H$_{16}^+$);
15\% coronene cation (C$_{24}$H$_{12}^+$); 7\% dicoronylene cation
(C$_{48}$H$_{20}^+$); 7\% benzo[b]fluoranthene cation
(C$_{20}$H$_{12}^+$); 7\% benzo[k]fl uoranthene cation
(C$_{20}$H$_{12}^+$); 7\% neutral naphthalene (C$_{10}$H$_8$); 4\%
naphthalene cation (C$_{10}$H$_{8}^+$); 4\% phenanthrene cation
(C$_{14}$H$_{10}^+$); 4\% chrysene cation (C$_{18}$H$_{12}^+$); 4\%
tetracene cation (C$_{18}$H$_{12}^+$).}.  The following constraints
and conclusions regarding the emitting astronomical PAH population can
be drawn from this comparison.  First, the overall agreement between
the astronomical emission spectrum and the simple average PAH spectrum
lend general support to the PAH model.  Second, the peak of the
dominant band in the CC stretching region for both small and large
pure PAHs falls longwards of the observed 6.2 \mum\, position,
reiterating the need to alter this fundamental vibration in some way.
The substitution of nitrogen for some of the carbon atoms satisfies
this constraint.  Third, the spectra shown here suggest that the
overall profile of the 7.7 \mum\, feature, with components at 7.6 and
7.8 \mum, may be accommodated by emission from a PAH population that
includes large and small PAHs.  The discussion above also suggests
that, in addition to larger PAHs contributing strongly to the 7.8
\mum\, component (dominant in Class B profiles), PAH anions contribute
more to the longer wavelengths than do PAH cations, consistent with
the spatial variation of the 7.8 \mum\, component in NGC~2023
\citep{Bregman:05}.  Fourth, a prominent peak near 8.6 \mum\, appears
only in the spectra of larger PAHs, suggesting that the relative
intensity of the 8.6 \mum\, astronomical band to the other
astronomical bands can be taken as an indicator of the relative
amounts of large to small PAHs in the emitting population.  For
example, the 8.6 \mum\, feature is barely evident on the wing of a
strong 7.7 \mum\, feature in some objects (e.g. IRAS07027-7934,
IRAS21190+5140), while it is prominent in others such as the objects
shown here, and it can be as intense at the 7.7 \mum\, band itself in
others (e.g. MWC~922).  In addition, large PAH anions are most likely
responsible for the red-shifted 8.6 \mum\, feature.

\subsection{The CH Out-of-Plane Bending Vibrations (9 to 15 \mum).}
\label{astro_112}

The astronomical emission bands arising from the CH$_{oop}$ bending
vibrations were discussed in detail by \citet{Hony:oops:01}.  Their
analysis was based on a detailed experimental study of 20 PAHs which
sampled the different edge structures possible and which range in size
from C$_{10}$H$_8$ to C$_{32}$H$_{14}$
\citep{Hudgins:tracesionezedpahs:99}.  Here we extend this analysis
to larger PAHs and assess the earlier conclusions.

The PAH bands in the CH$_{oop}$ region are used to gain insight into the
relative number of solo, duo, trio, and quartet hydrogens on
astronomical PAHs and perhaps provide some insight into their charge
distribution.  Indeed, the relative intensities of these bands have
been used to determine the relative amounts of the different types of
peripheral CH groups present in the emitting astronomical PAH
population and constrain astronomical PAH molecular structures.  In
addition, comparing the relative intensities of the CH$_{oop}$ bands with
the PAH emission features at shorter wavelengths and searching for
interband correlations has been used to further constrain
structures. This analysis has led to the general conclusion that the
astronomical bands in the 10.5 to 15 \mum\, region are dominated by
emission from large neutral PAHs and small aromatic grains whereas the
emission in the 5 to 9 \mum\, region is dominated by charged PAHs.

Figure \ref{fig_average}c shows the average spectrum of the neutral,
cation, and anion forms of the PAHs considered here in the CH$_{oop}$
region.  Taking the roughly 15 cm$^{-1}$ redshift ($\sim$0.2 \mum\, at
11 \mum) expected for emission into account, this figure shows that
the previous assignment of the 11.2 \mum\, astronomical emission band
to neutral PAHs and the 11.0 \mum\, astronomical band to PAH cations
\citep{Hudgins:tracesionezedpahs:99, Hony:oops:01} holds for these
larger species as well.

All the PAHs in this sample contain only solo and duo hydrogens.  As
discussed in Sect. \ref{lab_112}, the solo modes for these large,
compact, symmetric PAHs fall in the wavelength range expected for the
CH$_{oop}$ bend.  However, this isn't the case for the duo modes. Coupling
between the duo and solo CH$_{oop}$ modes splits the duo mode into two
bands (Figure \ref{fig_oops_D}).  The longer wavelength component of
the duo bands falls in the region traditionally assigned to trio
bands.

\begin{figure}
      \centering \includegraphics[width=.45\textwidth]{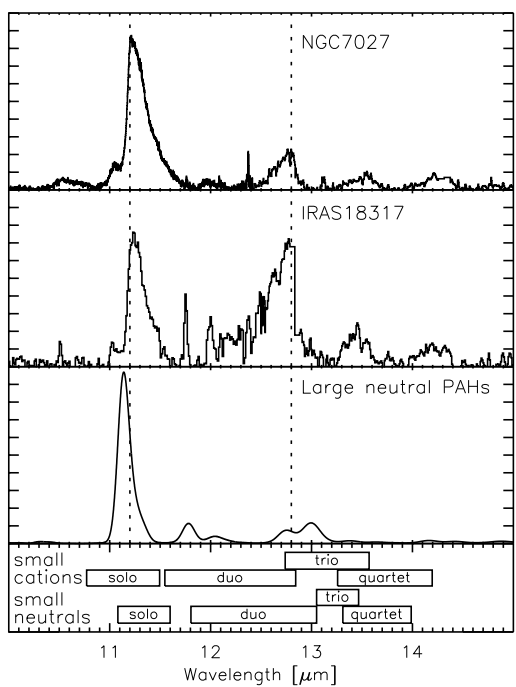}
\caption{Astronomical PAH emission spectra from 10 to 15 $\mu$m
    compared with the average spectrum of neutral, very large
    PAHs. Shown are the continuum subtracted IR spectrum of the
    planetary nebula NGC~7027 {\bf (top)}; the continuum subtracted IR
    spectrum of the HII region IRAS~18317-0757 {\bf (second)}; the
    average spectrum of the neutral large PAHs considered in this
    paper {\bf (third)}; and the ranges for the out-of-plane bending
    modes for small PAHs (C-atoms $\le$ 32) \citep{Hony:oops:01}{\bf
    (bottom)}. The data in the lower two frames have been redshifted by
    15 cm$^{-1}$ to correct for the wavelength shift between the
    absorption and emission process (see text).  }
\label{fig_oops}
\end{figure}

Figure \ref{fig_oops} compares the average spectrum of the neutral
PAHs considered here to the spectra of NGC~7027 and IRAS~18317-0757
from \citet{Hony:oops:01}.  These two object span the entire range in
observed 11.2/12.7 PAH intensity ratios as observed by
\citet{Hony:oops:01}. The computed spectrum is redshifted by
15 cm$^{-1}$ to account for the emission process.  The absence of bands
in the computed spectrum at 11.0, 13.5, and 14.3 \mum, the position of
several astronomical features in Figure 9, is to be expected.  These are
attributed to PAH cations, and CH$_{oop}$ bends from quartet and quintet
hydrogens respectively, species and structures which are not present
in the PAHs which produce the computed spectrum of large neutral PAHs.  The
good agreement between the strong 11.2 \mum\, astronomical
feature and the average of the computed spectra is in complete
agreement with it's assignment to large, neutral PAHs.

The previously unrecognized component of the duo modes near 13 \mum\,
overlaps the astronomical emission feature at 12.8 \mum\, which is
normally attributed to trio CH$_{oop}$ modes.  This duo band impacts the
interpretation of the astronomical PAH emission spectrum in two ways.
First, the number of trio hydrogens deduced with respect to solo and
duo hydrogens on the emitting astronomical PAH population must be
reduced.  The spectra shown in Figures \ref{fig_oops} and
\ref{fig_average}c suggest that an important fraction of the intensity
of the 12.8 \mum\, astronomical feature can arise from the CH$_{oop}$
bending vibrations from duo H's.  For NGC~7027, this may be as high as
50\%.  This, in turn, implies that the structures \citet{Hony:oops:01}
deduced as being most important in these objects and others with
spectra between 9 - 15 \mum\, which show clear evidence for the 12
\mum\, duo H band should be modified to more compact and symmetric
forms similar to those shown in Figure 1 here.  Second, this overlap
between the duo and trio bands in neutral PAHs may resolve a
longstanding puzzle regarding the origin of the blue shading on the
12.8 \mum\, astronomical feature.  The lower frame in Figure
\ref{fig_oops} shows that the CH$_{oop}$ trio CH bands for small neutral
PAHs fall in a narrow band, from 13.0 to 13.5 \mum\, while the CH$_{oop}$
duo bands for the neutral PAHs considered here range from 12.5 to 13.2
\mum. Figure \ref{fig_average}c shows this holds for the charged forms
of these PAHs as well.  Overlap between this duo band with the trio
feature could produce the observed, blue-shaded profile.

Keep in mind that this study involves only compact, symmetric PAHs
that contain only solo and duo hydrogens and that using lower symmetry
species will allow more types of hydrogens leading to additional
couplings.  These results suggest that caution is warranted when
constraining the number and nature of adjacent hydrogens and implied
structures using the out-of-plane bending modes alone.  This issue is
considered further in paper II which treats the spectra of less
symmetric and less compact very large PAHs.

\section{Conclusions}
\label{con}

The mid-IR spectra of seven very large, compact, symmetric PAHs with
formulae \stb, \stc, \std, \ste, \stf, \stg, and \sth\, have been determined
computationally using Density Functional Theory (DFT).  Previous to
this study, the mid-IR spectroscopic properties of PAHs was
essentially limited to species containing about 50 or fewer carbon
atoms.  These data provide new insight into the effect of size,
structure and charge on PAH IR spectra and enhances our understanding
of the PAH populations that contribute to the observed astronomical
spectra.  The main conclusions regarding the principle astronomical
emission features follow.

{\it The 3.29 \mum\, Feature:} The strongest feature in the CH
stretching region near 3.3 \mum\, is similar for all PAH species
considered in this work.  The peak position of this band in these
compact, very large PAHs falls within the regions expected based on
the spectroscopy of smaller PAHs, but covers a narrower range.  While
clearly ruling out an origin in PAH cations, these spectra suggest
that the astronomical A$_{3.3}$ and B$_{3.3}$ bands reflect the
relative contribution of PAH neutrals and anions.

{\it The 5 - 9 \mum\, Features:} Consistent with earlier work on smaller
PAHs, the intensities of the major features between 5 and 9 \mum\, in
these larger PAHs are enhanced by an order of magnitude or more upon
ionization.

The dominant bands in the CC stretching region between 6.1 and 6.4
\mum\, show less variation in the spectra of the anions than in the
spectra of the cations and they are slightly weaker relative to the
bands between 7 and 9 \mum.  The average band in the cation and anion
spectra peaks slightly redward of 6.3 \mum, far to the red of the
astronomical 6.2 \mum\, band.  Thus, as with small PAHs, the CC
stretch in large PAHs comprised of only C and H cannot reproduce the
peak position of the astronomical feature, reinforcing our previous
suggestion that astronomical PAHs contain nitrogen
\citep[PANHs,][]{Peeters:prof6:02, Hudgins:05}.

The largest PAH cations and anions considered have strong bands close
to 7.7 \mum, the position of the strongest astronomical emission
feature.  The strong feature produced by the cations and anions
overlap to a large extent, but the average anion peak falls about 0.1
\mum\, to the red of the average cation peak.  These data suggest that
the astronomical 7.7 \mum\, band is produced by overlapping bands from
a mixture of small and large PAH cations and anions, with small PAHs
contributing more to the 7.6 \mum\, component and large PAHs more to
the 7.8 \mum\, component. This implies that the variation in the peak
position of the 7.8 \mum\, feature may be related to the variation
in relative amounts of large PAH cations and anions since negatively
charged large PAHs emit at slightly longer wavelenths than do the PAH cations.

A band near 8.5 \mum\, grows in strength and shifts slightly to the
red as PAH size increases for both anions and cations.  The bands from
both the large PAH anions and cations overlap, with the anions
contributing more strongly at slightly longer wavelengths.
\citet{Peeters:prof6:02} found a correlation between the peak position
of the astronomical 7.7 \mum\, feature and that of the 8.6 \mum\,
band.  The inability of small PAHs to reproduce this correlation
suggests that the astronomical 7.8 \mum\, component and the 8.6 \mum\,
feature originate primarily in large, cationic and anionic PAHs, with
the specific peak position and profile reflecting the particular
cation to anion concentration ratio in any given object.

{\it The 10 - 15 \mum\, Features:} The prominent features that span
this region correspond to CH$_{oop}$ bending vibrations.  The large
PAHs in this sample contain only solo and duo hydrogens.  The solo
modes for these large, compact, symmetric PAHs fall in the wavelength
range expected for the solo CH$_{oop}$ bend.  However, this isn't the
case for the duo CH$_{oop}$ modes.  The duo and solo CH$_{oop}$
vibrations couple in these large, compact PAHs, causing a splitting of
the duo band.  One component remains close to the traditional duo band
position while the other shifts to about 13 \mum, the region
traditionally assigned to trio bands.

The excellent agreement between the strong 11.2 \mum\, astronomical
feature and the average of the computed spectra for neutral PAHs
presented here lends strong support to its assignment to large,
neutral astronomical PAHs.  However, the previously unrecognized
component of the duo modes near $\sim$12.8 \mum\, overlaps the
astronomical emission feature at 12.8 \mum\, which is normally
attributed to trio CH$_{oop}$ modes.  This overlap impacts the
interpretation of the astronomical PAH emission spectrum in two ways.
The number of trio hydrogens deduced with respect to solo and duo
hydrogens on the emitting astronomical PAH population must be reduced,
in some cases, by as much as a factor of two, implying that the PAH
structures \citet{Hony:oops:01} previously deduced should be modified
to favor more compact and symmetric forms.  Second, the overlap
between the duo and trio bands in neutral PAHs may produce the
puzzling, blue-shaded profile of the 12.8 \mum\, feature that is
commonly observed.

The spectra of comparable large, but less symmetric PAHs will be
presented and discussed in a forthcoming publication.

\acknowledgements We very gratefully acknowledge sustained support
from NASA's Long Term Space Astrophysics and Astrobiology Programs,
and the Spitzer Space Telescope Archival and General Observer Program.

\end{document}